\documentclass[iop, apj, natbib, numberedappendix]{emulateapj}
\usepackage{apjfonts} 
\usepackage[breaklinks,colorlinks,citecolor=blue,linkcolor=magenta]{hyperref}
\usepackage[colorlinks,citecolor=blue,linkcolor=magenta]{hyperref}
\usepackage{graphicx}
\usepackage{amsmath}
\usepackage{lmodern}
\usepackage{amssymb}
\usepackage{color}
\usepackage{multirow}
\usepackage{rotating}
\usepackage{booktabs}
\usepackage{times}
\usepackage{natbib}

\newcommand{\Msun}{\ensuremath{M_{\odot}}}

\newcommand{\ar}{\arcsec}
\newcommand{\zp}{\ensuremath{z^\prime}}
\newcommand{\ip}{\ensuremath{i^\prime}}

\newcommand{\nbfot}{\textit{NB503}}
\newcommand{\nbfso}{\textit{NB570}}
\newcommand{\nbest}{\textit{NB816}}

\newcommand{\nbnto}{\textit{NB921}}

\newcommand{\lya}{Ly$\alpha$}
\newcommand{\ha}{H$\alpha$}
\newcommand{\hb}{H$\beta$}
\newcommand{\oii}{[O\,{\sc ii}]}	
\newcommand{\oiii}{[O\,{\sc iii}]}
\newcommand{\nev}{[Ne\,{\sc v}]}
\newcommand{\mgii}{Mg\,{\sc ii}}
\newcommand{\feii}{Fe\,{\sc ii}}
\newcommand{\feiis}{Fe\,{\sc ii}$^{\ast}$}

\newcommand{\hii}{H\,{\sc ii}}

\newcommand{\oiib}{O\,{\sc ii}B}	
\newcommand{\oiiib}{O\,{\sc iii}B}	
\newcommand{\hab}{H{\sc A}B}	

\newcommand{\rz}{\textit{Rz}}
\newcommand{\bpv}{\textit{BV}}
\newcommand{\br}{\textit{BR}}

\newcommand{\um}{$\micron$}
\newcommand{\kms}{km\,s$^{-1}$}
\newcommand{\ergs}{erg\,s$^{-1}$}

\newcommand{\ergscm}{erg\,s$^{-1}$\,cm$^{-2}$}

\slugcomment{Accepted for publication in ApJ}

\shorttitle{Blob Evolution up to $\MakeLowercase{z}=1.5$}
\shortauthors{Yuma et al.}

\begin{document}

\title{Systematic Survey for \oii, \oiii, and \ha\ Blobs at $\MakeLowercase{z}=0.1-1.5$: \\
The Implication for Evolution of Galactic-Scale Outflow
}
\author{Suraphong Yuma$^{1}$}
\author{Masami Ouchi$^{2}$}
\author{Alyssa B. Drake$^{3}$}
\author{Seiji Fujimoto$^{2,4}$}
\author{Takashi Kojima$^{2,5}$}
\author{Yuma Sugahara$^{2,5}$}

\affil{$^1$Department of Physics, Faculty of Science, Mahidol University, Bangkok 10400, Thailand; suraphong.yum@mahidol.ac.th}
\affil{$^2$Institute for Cosmic Ray Research, The University of Tokyo, Kashiwa-no-ha, Kashiwa 277-8582, Japan}
\affil{$^3$CRAL, Observatoire de Lyon, Universit\'e Lyon 1, 9 avenue Charles Andr\'e, 69561 Saint Genes Laval Cedex, France}
\affil{$^4$Department of Astronomy, University of Tokyo, Hongo, Bunkyo-ku, Tokyo 113-0033, Japan}
\affil{$^5$Department of Physics, University of Tokyo, Hongo, Bunkyo-ku, Tokyo 113-0033, Japan}


\begin{abstract}

We conduct a systematic search for galaxies at $z=0.1-1.5$ 
with \oii$\lambda3727$, \oiii$\lambda5007$, or 
\ha$\lambda6563$ emission lines extended over at least 30 kpc by using 
deep narrowband and broadband imaging in Subaru-{\it XMM} Deep Survey (SXDS) field. 
These extended emission-line galaxies are dubbed \oii, \oiii, or \ha\ blobs. 
Based on a new selection method that securely select extended emission-line galaxies, 
we find 77 blobs at $z=0.40-1.46$ with the isophotal area of emission lines down to 
$1.2\times10^{-18}$ \ergscm kpc$^{-2}$. 
Four of them are spectroscopically confirmed to be \oiii\ blobs at $z=0.83$. 
We identify AGN activities in 8 blobs with X-ray and radio data, and 
find that the fraction of AGN contribution increases with increasing isophotal area of the extended emission. 
With the Kolmogorov-Smirnov (KS) and Anderson-Darling tests, 
we confirm that the stellar-mass distributions of \ha\ and \oii\ blobs are not drawn from those of 
the emitters at the $>90$\% confidence level in that \ha\ and \oii\ blobs are located at the massive end 
of the distributions, 
but cannot reject null hypothesis of being the same distributions in terms of the specific star formation rates. 
It is suggested that galactic-scale outflows tend to be more prominent in 
more massive star-forming galaxies. 
Exploiting our sample homogeneously selected over the large area, 
we derive the number densities of blobs at each epoch. 
The number densities of blobs decrease drastically with redshifts 
at the rate that is larger than that of the decrease of cosmic star formation densities. 

\end{abstract}

\keywords{galaxies: high redshift --- galaxies: evolution --- galaxies: formation}

\section{Introduction}\label{sec:intro}


It has long been known that galaxies in the local universe can be divided into categories according to 
the so-called Hubble sequence. 
Main populations are spiral/disk galaxies with active star-forming activity 
and elliptical galaxies that rarely form new stars and evolve passively. 
Astronomers suggested that progenitors of elliptical galaxies 
should somehow stop forming new stars and start evolving passively at some epoch 
around $z\sim1$ \citep[e.g., ][]{trujillo04, mcintosh05}. 
The cessation of star formation in galaxies occurs when gas in those galaxies is used up 
and/or galactic-scale outflow mechanism 
expels cool gas out of galaxies 
at the rate larger than the accretion of inflowing gas \citep[e.g.,][]{lilly13}. 
Feedback from active galactic nuclei (AGN) and supernova explosions has been introduced to explain 
the gas outflow and subsequently the suppression of star formation. 
It is a key process in turning star-forming galaxies into 
passively evolving massive ellipticals. 
In a theoretical framework, 
strong radiation from an accretion disk and radio jet of an AGN at the center of a galaxy 
prevents a cooling flow of gas, leading to 
star-formation quenching in massive galaxies at the massive end of the stellar mass function \citep[e.g., ][]{somerville08}. 
On the other hand, supernova and stellar winds heat up cold gas in the interstellar medium (ISM) 
that is expelled by outflow and suspends the star formation 
in low-mass galaxies \citep[e.g., ][]{benson03}. 

To date, outflows have been extensively studied in various 
types of galaxies; e.g., Ultra Luminous Infrared Galaxies \citep[ULIRGs, ][]{heckman90, martin05, soto12}, 
Sub-Millimeter Galaxies \citep[SMGs, ][]{alexander10},  
and star-forming galaxies \citep{steidel10, martin12, kornei12, erb12, bradshaw13}. 
Recently, \cite{cheung16} studied the outflow in AGN hosting quiescent galaxies 
and showed that an active SMBH even in low-luminosity AGNs can provide sufficient energy to drive the outflow. 
However, study of gas outflow at a large scale, which is closely related to the 
quenching process of star formation and subsequently to the evolution of elliptical galaxies, 
is still incomplete, especially for non-AGN galaxies.

The large-scale outflow of gas takes place in a relatively short timescale; therefore, it is difficult for observations 
to pinpoint a galaxy in the middle of such a major process that subsequently stops star-forming activity. 
\citeauthor{yuma13} (2013, hereafter Y13) 
introduced the first systematic search for galaxies with spatially extended 
\oii$\lambda\lambda3726, 3729$ emission (dubbed ``\oii\ blobs" or ``\oiib s"). 
The idea of their method is to select the galaxies with 
the strong \oii\ emission line that is redshifted and falls into the narrowband filter and 
examine the spatial extension of the \oii\ emission line from the narrowband image after subtracting 
contribution of the stellar continuum. 
With deep narrowband imaging covering a survey volume of $1.9\times10^5$ Mpc$^3$ 
in the Subaru XMM Deep Survey (SXDS), 
Y13 discovered a giant \oii\ blob, named \oiib1, with a spatial extent of \oii\ emission 
over 75 kpc and identified a total of 12 \oiib s with $>30$-kpc extension at $z\sim1.2$. 
\oiib1 is identified as an AGN with the $6\sigma$ detection of the high-excitation \nev$\lambda3426$ emission line, 
while the others are star-forming galaxies with no AGN activity. 
According to the surface brightness profiles, 
the oxygen emission of these \oiib s is extended beyond the stellar component 
and is thought to be due to hot metal-rich gas outflowing from 
galaxies rather than pristine gas inflowing from metal-poor 
intergalactic medium \citep[IGM; e.g.,][]{aguirre08, fumagalli11}. 
The outflow hypothesis is confirmed by the spectroscopic analysis of 
three \oiib s \citep[Y13; ][]{harikane14}. 
The spectrum of \oiib 1 with the largest extent of \oii\ emission line show the blueshifted 
\feii $\lambda2587$ absorption and \feiis $\lambda2613$ emission lines with 
the outflow velocities of $500-600$ \kms. 
The outflow in the other two \oiib s is examined from the blueshifted \mgii $\lambda2796, 2804$ and 
\feii $\lambda2587$ absorption lines. The derived outflow velocities range from 80 \kms\ to 260 \kms, 
which is consistent with those of normal star-forming galaxies. 
Furthermore, \cite{harikane14} used line ratios to examine the energy source of another \oiib\ at $z\sim1.2$ 
by using the blue diagram \citep[\oiii/\hb\ versus \oii/\hb; ][]{lamareille04, lamareille10} 
and found that the outflow is powered by both star formation and the AGN.  
Searching for galaxies with extended oxygen emission lines provides 
a systematic sample of galaxies showing metal-rich large-scale outflow both with and without AGNs. 
Although extended oxygen emission is already seen in some AGNs \citep[e.g., ][]{nesvadba08}, 
this method efficiently provides samples of large-scale outflows in non-AGN galaxies for the first time. 

As a systematic search, Y13 are able to determine the number density of the \oiib s at $z\sim1.2$. 
It is $\sim5\times10^{-6}$ Mpc$^{-3}$ for OIIB1-type giant blobs with an AGN. 
The number density is comparable to that of the AGNs with outflow at the similar redshift 
\citep{barger05, barger_cowie05}. 
The number density of all \oiib s at $z\sim1.2$ including \oiib 1 is $6.3\times 10^{-5}$ Mpc$^{-3}$. 
By comparing the \oiib\ number density with those of star-forming galaxies at the similar redshift, 
it is implied that 3\% of star-forming galaxies at $z\sim1$ are quenching 
the star formation through an outflow involving extended \oii\ emission. 

Although this method is successful in systematically selecting galaxies with large-scale outflows 
by using only the imaging data, Y13 only focused their study at one specific epoch of the universe, 
i.e., at $z\sim1.2$. 
In order to understand the wider picture of gas outflow, 
AGN/stellar feedback, and eventually the galaxy evolution, 
we push our study toward other epochs of the universe.
Furthermore, we expand our search by applying 
the idea of the extended \oii\ emission line indicating an outflow, 
to other optical emission lines such as \oiii$\lambda5007$ and \ha$\lambda6563$,  
which are the strong emission lines in star-forming galaxies and AGNs. 
Although the \ha\ emission line is not a metal emission line 
such as \oii\ or \oiii\ emission, extended \ha\ emission seen 
in local starbursts with strong outflow like M82 suggests that spatially extended \ha\ emission 
line can also be used as an indicator for galactic-scale outflow. 
In fact, \cite{lin17} recently discovered a nearby galaxy with extended \ha\ emission, 
which is plausibly caused by an AGN outflow. 
This suggests that the spatial extension of the \ha\ emission line can be primarily used to 
indicate the large-scale outflow from the galaxy. 
With currently available narrowband and broadband imaging data from SXDS, 
we conduct a systematic study of large-scale outflowing galaxies selected with 
extended \oii, \oiii, and \ha\ emission lines and their 
evolution over the past 9 billion years from $z\sim1.5$ to the present-day universe. 

This paper is organized as follows. 
Section \ref{sec:data} describes the imaging data and selection methods including 
a new process we develop to efficiently select galaxies with extended emission lines. 
The results of blobs with different emission lines including spectroscopic confirmation 
are explained in section \ref{sec:results}. 
In section \ref{sec:discuss}, we discuss in detail about the AGN contribution, stellar properties, 
cumulative luminosity function, and the evolution of blobs along redshifts. 
Finally, we summarize all results in section \ref{sec:summary}. 
Throughout this paper, we adopt the standard $\Lambda$CDM cosmology with 
$H_0=70$ \kms\,Mpc${-1}$, $\Omega_m=0.3$, and $\Omega_\lambda=0.7$. 
All magnitudes are given in the AB system \citep{oke83}. 


\section{Data and Sample Selection}\label{sec:data}

\subsection{Imaging data}

\begin{figure}
\centering
\includegraphics[width=0.45\textwidth]{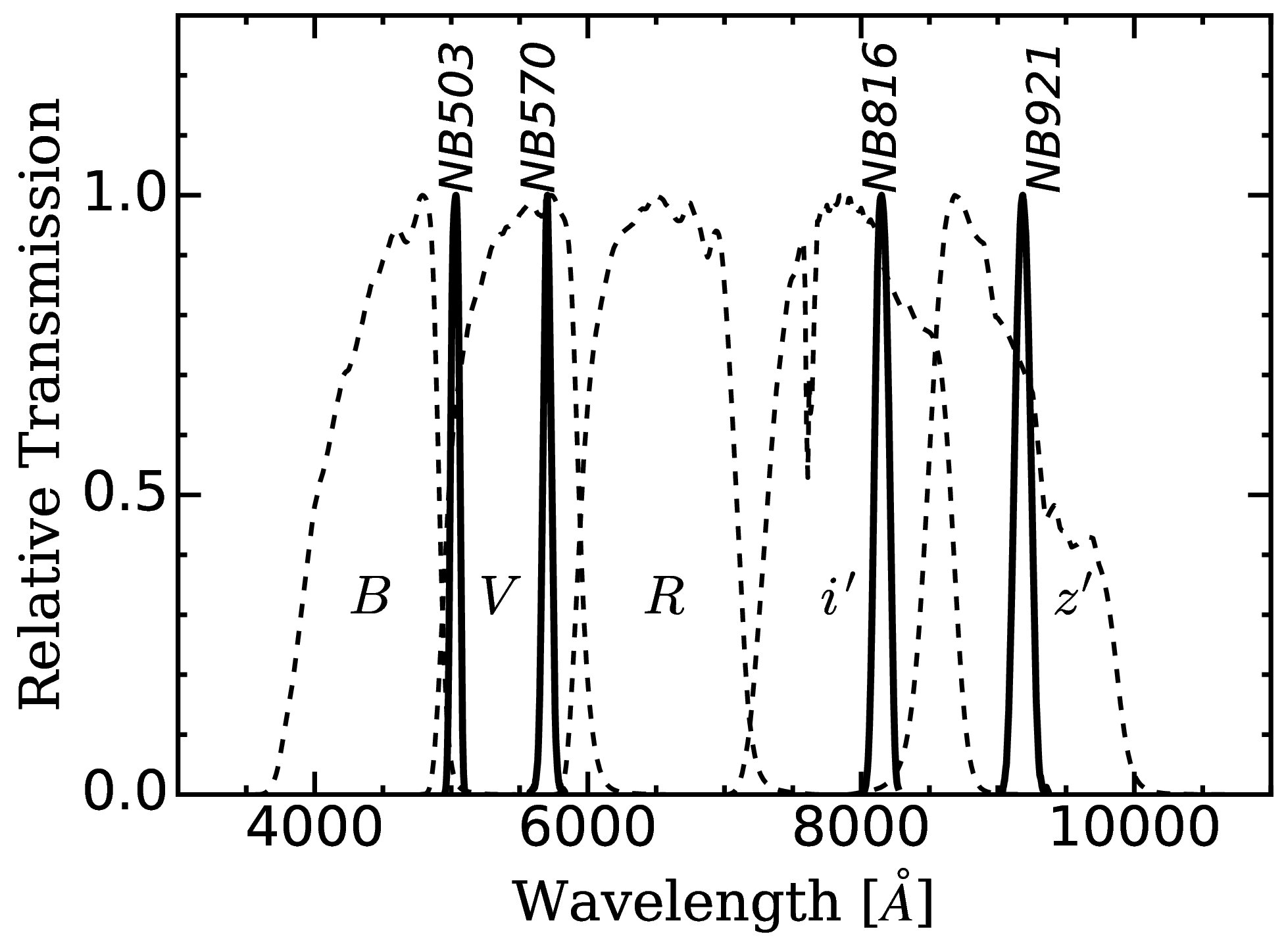}
\caption{
	 Peak-normalized transmission curves of Subaru/Suprime-Cam broadband 
	 $BVR$\ip\zp\ and narrowband \nbfot, \nbfso, \nbest, and \nbnto\ fillters.  
}
\label{fig_filters}
\end{figure}

Optical data in the SXDS field are obtained with Suprime-cam on the Subaru Telescope 
in four narrowband \citep[\nbfot, \nbfso, \nbest,  and \nbnto ; ]
[]{ouchi08, ouchi09} and 
five broadband filters \citep[$B$, $V$, $R$, \ip, and \zp ; ][]{furusawa08}. 
All data cover an area of 1.3 deg$^2$ in five overlapping pointings of Suprime-Cam.
The combination of these filters is used to select emitters at all redshifts. 
Figure \ref{fig_filters} shows transmission curves of the above filters. 
Near-infrared data in $J$, $H$, and $K$ 
bands are obtained from data release 8 (DR8) of 
UKIDSS ultra deep survey \citep[UDS; ][]{lawrence07} 
taken with WFCAM on the UK Infrared Telescope (UKIRT) 
and mid-infrared data in Infrared Array Camera (IRAC) channels 
1 (3.6\um) and 2 (4.5\um) from {\it Spitzer} UDS survey 
(SpUDS; PI: J. Dunlop). 
The final overlapping area of Subaru and 
other surveys is 0.63 deg$^2$ \citep{drake13}. 
Table \ref{tab_data} summarizes crucial information of 
all images used in this paper including central wavelength, 
full width at half maximum (FWHM), and $5\sigma$ limiting magnitude.

\begin{deluxetable*}{cccccc}
\tabletypesize{\footnotesize}
\tablewidth{0pt}
\tablecolumns{6}
\tablecaption{Summary of photometric data in all narrowband and broadband filters \\
from Subaru, CFHT, UKIRT, and {\it Spitzer} in SXDS field. \label{tab_data}}
\tablewidth{0pt}
\tablehead{
\multicolumn{1}{c}{Telescope/Instrument} &
\multicolumn{1}{c}{Filter} &
\multicolumn{1}{c}{Central Wavelength (\AA)} &
\multicolumn{1}{c}{FWHM (\AA)} & 
\multicolumn{1}{c}{Limiting Magnitude\tablenotemark{a}} & 
\multicolumn{1}{c}{Reference}
}
\startdata
	  Subaru/Suprime-Cam & {\it NB503} & 5029 & 73 & 24.9 & \cite{ouchi08}\\
	  Subaru/Suprime-Cam & {\it NB570} & 5703 & 68 & 24.5 & \cite{ouchi08}\\
	  Subaru/Suprime-Cam & {\it NB816} & 8150 & 119 & 25.7 & \cite{ouchi08}\\
	  Subaru/Suprime-Cam & {\it NB921} & 9196 & 132 & 25.4 & \cite{ouchi09}\\
	  Subaru/Suprime-Cam & $B$ & 4473 & 1079 & 27.4 & \cite{furusawa08}\\
	  Subaru/Suprime-Cam & $V$ & 5482 & 984 & 27.1 & \cite{furusawa08}\\
	  Subaru/Suprime-Cam & $R$ & 6531 & 1160 & 26.9 & \cite{furusawa08}\\
	  Subaru/Suprime-Cam & \ip & 7695 & 1543 & 26.7 & \cite{furusawa08}\\
	  Subaru/Suprime-Cam & \zp & 9149 & 1384 & 26.0 & \cite{furusawa08}\\
	  UKIRT/WFCAM & $J$ & 12500 & 1570 & 25.0 & \cite{lawrence07}\\
	  UKIRT/WFCAM & $H$ & 16500 & 2910 & 24.3 & \cite{lawrence07}\\
	  UKIRT/WFCAM & $K$ & 22000 & 3530 & 24.6 & \cite{lawrence07}\\
	  {\it Spitzer}/IRAC & ch1 & 35500 & 7411 & 23.6 & SpUDS (PI: J. Dunlop)\\
 	  {\it Spitzer}/IRAC & ch2 & 44900 & 10072 & 22.9 & SpUDS (PI: J. Dunlop)

\enddata
\tablenotetext{a}{Limiting magnitudes at $5\sigma$ in 2.0\ar\ diameter aperture.}
\end{deluxetable*}

\subsection{Spectroscopic Data}\label{subsec:specdata}

We carried out spectroscopic follow-up observations for four \oiii blob candidates 
with Subaru/Faint Object Camera and Spectrograph \citep[FOCAS; ][]{kashikawa02} 
on October $22-23$, 2014 and December $2-4$, 2015. 
The sky was mostly clear, expect the first-half of October 22, 2014. 
The on-source exposure time ranges from 7,200 to 18,000 s.
The FOCAS observations were performed in MOS mode with VPH450 
grating and VPH850 grating with SO58 order-cut filter. 
The blaze wavelengths of VPH450 and VPH850 are 4000 \AA\ and 8000 \AA\ 
with the spectral ranges of $3800-5250$ \AA\ and $5800-10350$ \AA, respectively. 
With a slit width of 0.8\ar, the resolutions are $\lambda/\Delta\lambda\simeq1700$ and 750 
for VPH450 and VPH850 gratings, respectively.

\subsection{Sample selection}\label{subsec:sample}

\oii\ blobs or \oiib s at $z\sim1.2$ in Y13 
are originally selected as \oii\ emitters by \cite{drake13} and studied in \cite{drake15}. 
They isolated \oiib s at from the normal \oii\ emitters with no extended emission 
based on an emission-line image constructed by subtracting 
interpolated continuum from the narrowband image. 
The \oiib s are those with isophotal area measured down to $2\sigma$ 
limiting surface brightness fluctuation of $28.0$ mag arcsec$^{-2}$ 
or $1.2\times10^{-18}$ \ergscm\,arcsec$^{-2}$ in the emission-line image 
over 900 kpc$^2$ or approximately 30 kpc spatial extent. 
Visual verification was then performed to remove merging systems 
that can mimic the extended feature of emission. 

In this paper, we follow the above processes used in Y13 and strengthen 
the selection process by including quantitative analysis 
instead of using visual verification to 
determine whether or not the galaxies really show the extension of 
the emission line beyond their stellar component. 
Briefly speaking, we perform a careful search of blobs according to the following processes. 

\begin{itemize}
	\item We select emitters or galaxies with strong emission lines (Section \ref{select_emitters}). 
	Table \ref{tab_emitters_sxds} summarizes types of emission lines 
	detected in each narrowband filter and the corresponding redshift ranges. 
	
	\item Then we construct the image that contains only flux 
	from the emission line by using broadband and narrowband images 
	and select only emitters with a spatially extended profile (Section \ref{select_blobs}). 
	
	\item Bright point spread functions (PSFs) and merging systems 
	that can mimic the extended profile of blobs 
	are excluded by a new quantitative method (Section \ref{new_method}). 
\end{itemize}

\subsubsection{Emitter selection}\label{select_emitters}

We use the sample of \oii, \oiii, and \ha\ emitters 
detected in four narrowband filters provided by \cite{drake13}. 
Detailed procedures for selecting the emitters are described in the above paper. 
In short, 
they identified objects with potential emission lines by using the narrowband excess, 
i.e., {\it Cont-NB}, where {\it Cont} and {\it NB} are the magnitudes of objects in broadband 
and narrowband images, respectively. 
The objects showing the strong emission line are those with the narrowband excess 
above $3\sigma$ level relative to the sigma-clipped median {\it Cont-NB} 
of all objects in the images. 
It roughly corresponds to the observed equivalent width (EW$_{\rm obs}$) of $30$ \AA. 
The photometric redshifts are then used to categorize emitters into 
\oii, \oiii, or \ha\ emitters at corresponding redshifts. 
The numbers of emitters at redshifts ranging from $z\sim0.1$ to $z\sim1.5$
are listed in Table \ref{tab_emitters_sxds}.

\begin{deluxetable}{cccc}
\tabletypesize{\footnotesize}
\tablecolumns{4}
\tablecaption{Summary of emitters in the SXDS field at each redshift bin taken from \cite{drake13}. 
\label{tab_emitters_sxds}}
\tablewidth{0pt}
\tablehead{
\multicolumn{1}{c}{Narrowband filter} &
\multicolumn{1}{c}{Emission line} &
\multicolumn{1}{c}{Redshift range\tablenotemark{a}} &
\multicolumn{1}{c}{\# of emitters} \\
}
\startdata
	  {\it NB503} & \oii  & $0.10 < {\bf 0.35} < 0.50$ & 142\\\\
 
 	  {\it NB570} & \oiii & $0.00 < {\bf 0.14} < 0.30$ & 42 \\
	  {\it NB570} & \oii  & $0.30 < {\bf 0.53} < 0.70$ & 96 \\\\

	  {\it NB816} & \ha  & $0.00 < {\bf 0.25} < 0.35$ & 152\\
	  {\it NB816} & \oiii & $0.35 < {\bf 0.63} < 0.80$ & 985\\
	  {\it NB816} & \oii  & $0.80 < {\bf 1.19} < 1.50$ & 1013\\\\
	  
	  {\it NB921} & \ha  & $0.00 < {\bf 0.40} < 0.50$ & 279\\
	  {\it NB921} & \oiii  & $0.50 < {\bf 0.83} < 1.10$ & 930\\
	  {\it NB921} & \oii   & $1.10 < {\bf 1.46} < 1.90$ & 2204\\
	  
\enddata
\tablenotetext{a}{Redshift ranges are photometric redshifts that are used 
to select emitters at each redshift bin. }
\end{deluxetable}

\subsubsection{Blob selection}\label{select_blobs}

\begin{deluxetable}{cc}
\tabletypesize{\footnotesize}
\tablewidth{0pt}
\tablecolumns{2}
\tablecaption{Limiting Surface Magnitudes of Emission-Line images. 
\label{tab_emission_image}}
\tablehead{
\colhead{Emission-line image} &
\colhead{$2\sigma$ Limitting surface magnitude/flux}\\
 & \colhead{(AB mag\,arcsec$^{-2}$/\ergscm\,arcsec$^{-2}$})
}
\startdata
	  {\it NB503-BV}\tablenotemark{a} & 27.8/$2.4\times 10^{-18}$\\
 	  {\it NB570-BR}\tablenotemark{b} & 27.2/$3.0\times 10^{-18}$\\
	  {\it NB816-Rz}\tablenotemark{c} & 28.0/$1.2\times 10^{-18}$ \\	  
	  {\it NB921-\zp} & 28.1/$1.1\times 10^{-18}$
\enddata
\tablenotetext{a}{\bpv\ is defined as {\it BV $\equiv$ (B+V)/2}.}
\tablenotetext{b}{\br\ is defined as {\it BR $\equiv$ (B+2R)/3}.}
\tablenotetext{c}{\rz\ is defined as {\it Rz $\equiv$ (R+2\zp)/3}.}
\end{deluxetable}

We construct the emission-line image by subtracting the continuum 
from the narrowband images. 
The continuum is determined from the broadband image at approximately 
the same wavelength as the narrowband images. 
Before performing the image subtraction, 
we resemble the PSF sizes of the narrowband and broadband images 
by smoothing the image with a Gaussian kernel 
to match the other one with the largest PSF size 
\citep[typically $0.8-1.0$\ar; ][]{ouchi08}. 
The remaining flux in the subtracted narrowband images 
would be only from the emission line. 
Surface magnitude limits of the emission-line images are 
summarized in Table \ref{tab_emission_image}. 
The resulting images are then smoothed with 
the Gaussian kernel.

\begin{figure*}
	\centering
	\begin{tabular}{cc}
		\includegraphics[width=0.3\textwidth]{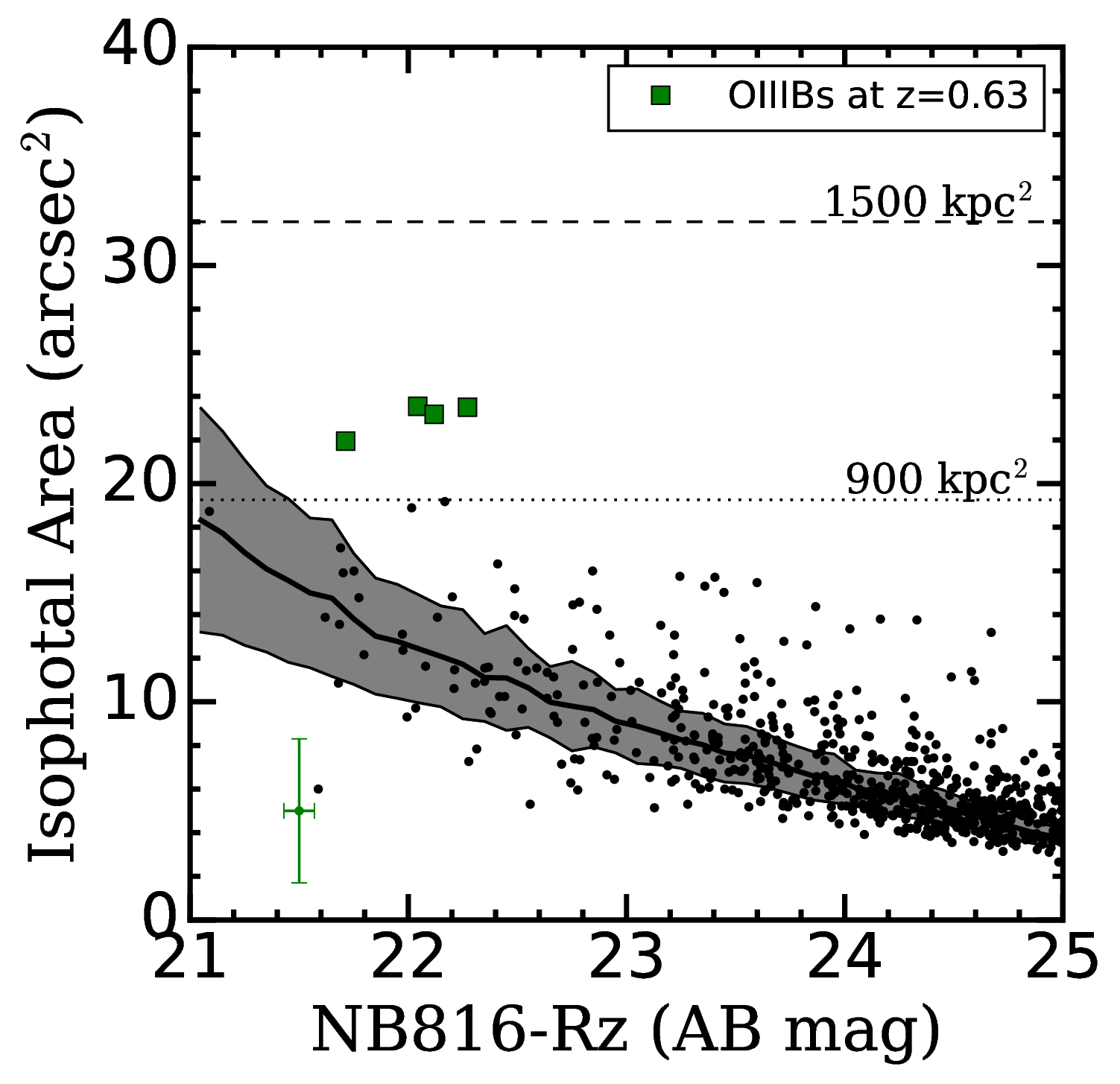} & 
		\includegraphics[width=0.3\textwidth]{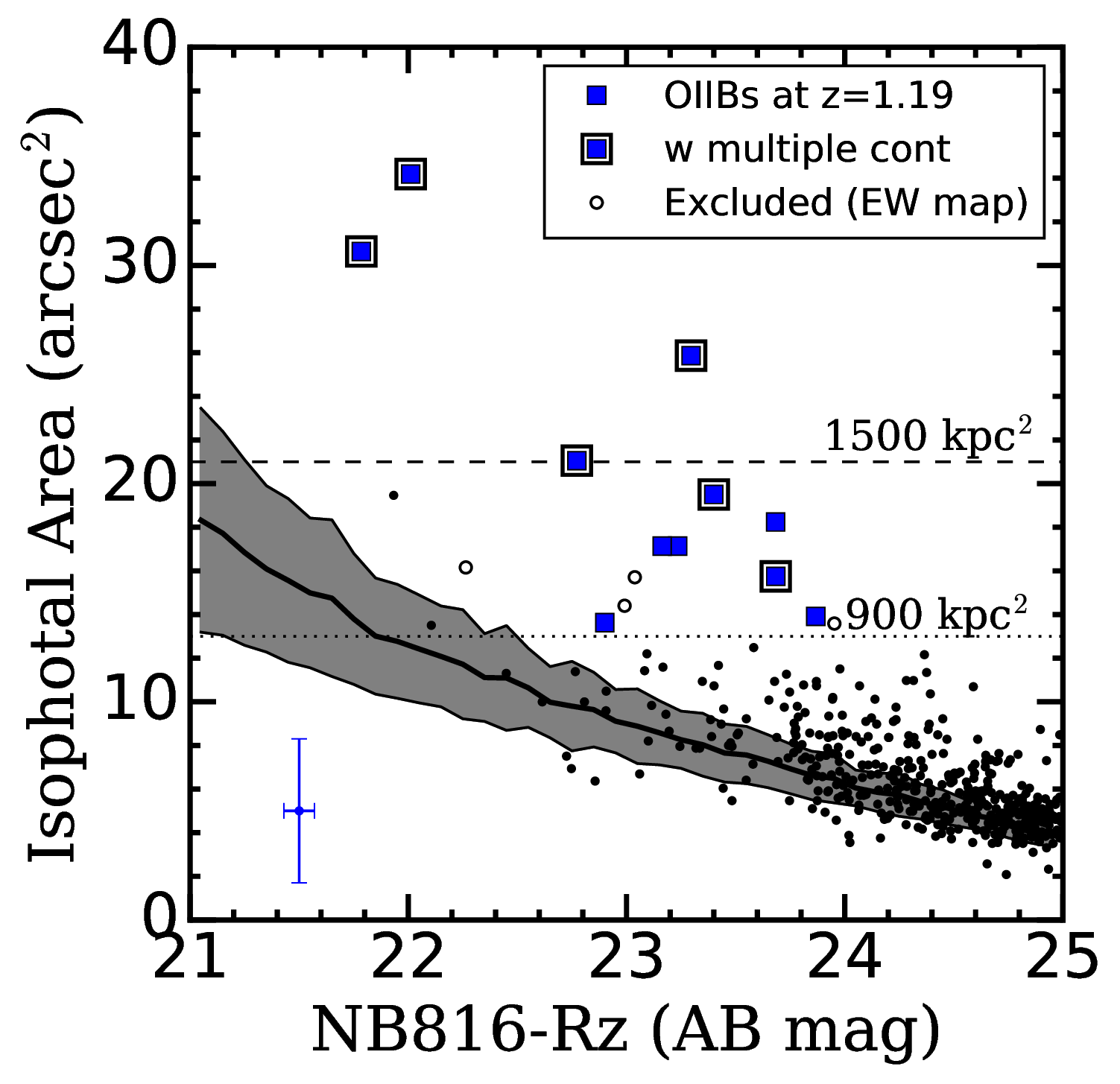} \\
	\end{tabular}
	\caption{Isophotal area-magnitude plot of the \nbest-\rz\ images for 
		\oiiib s at $z\sim0.63$ ({\it left}) and 
		\oiib s at $z\sim1.18$ ({\it right}). 
		Green and blue squares represent the \oiiib s and \oiib s 
		at respective redshifts. 
		Dotted symbols are all emitters at corresponding redshifts. 
		Those above the criteria of 900 kpc$^2$ are the emitters 
		the large isophotal area of which is caused by multiple continuum components. 
		They are excluded from the final blob sample by the PSF simulation 
		(Section \ref{new_method}). 
		Open circles are the emitters that are ruled out from the final blob sample 
		according to the EW-map criterion. 
		Dotted and dashed lines show isophotal criteria used to select 
		the blobs at 900 kpc$^2$ and 1500 kpc$^2$, respectively.   
		Blobs with multiple continuum components 
		are enclosed by black square symbols. 
		Error bars on the bottom left corner of the figures 
		indicate typical errors of magnitudes and isophotal area. 
		The grey regions show $1\sigma$ distribution of the simulated PSFs. 
	}
	\label{fig_iso_mag_nb816}
\end{figure*}

\begin{figure*}
	\centering
	\begin{tabular}{ccc}
		\includegraphics[width=0.31\textwidth]{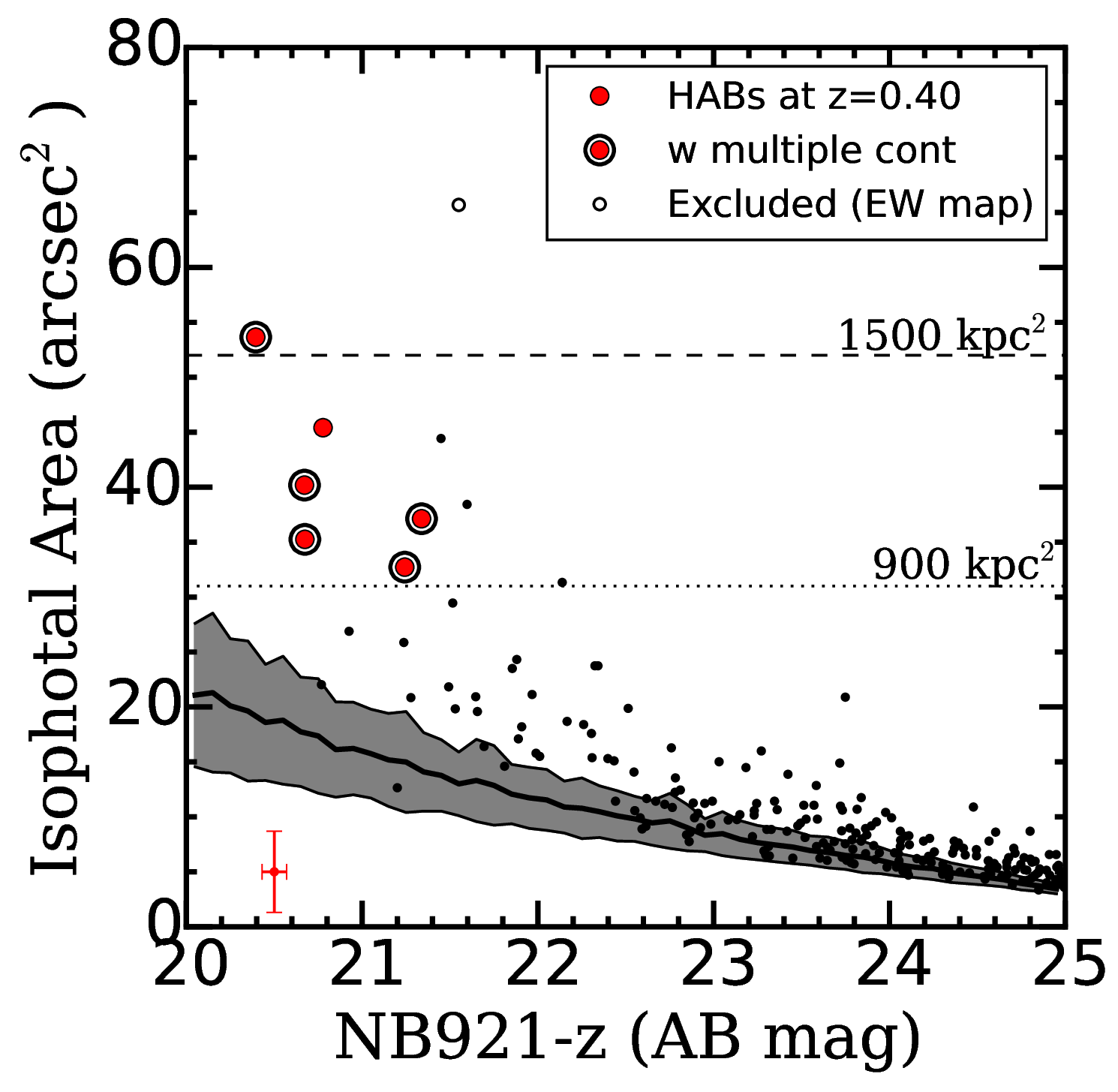} &
		\includegraphics[width=0.3\textwidth]{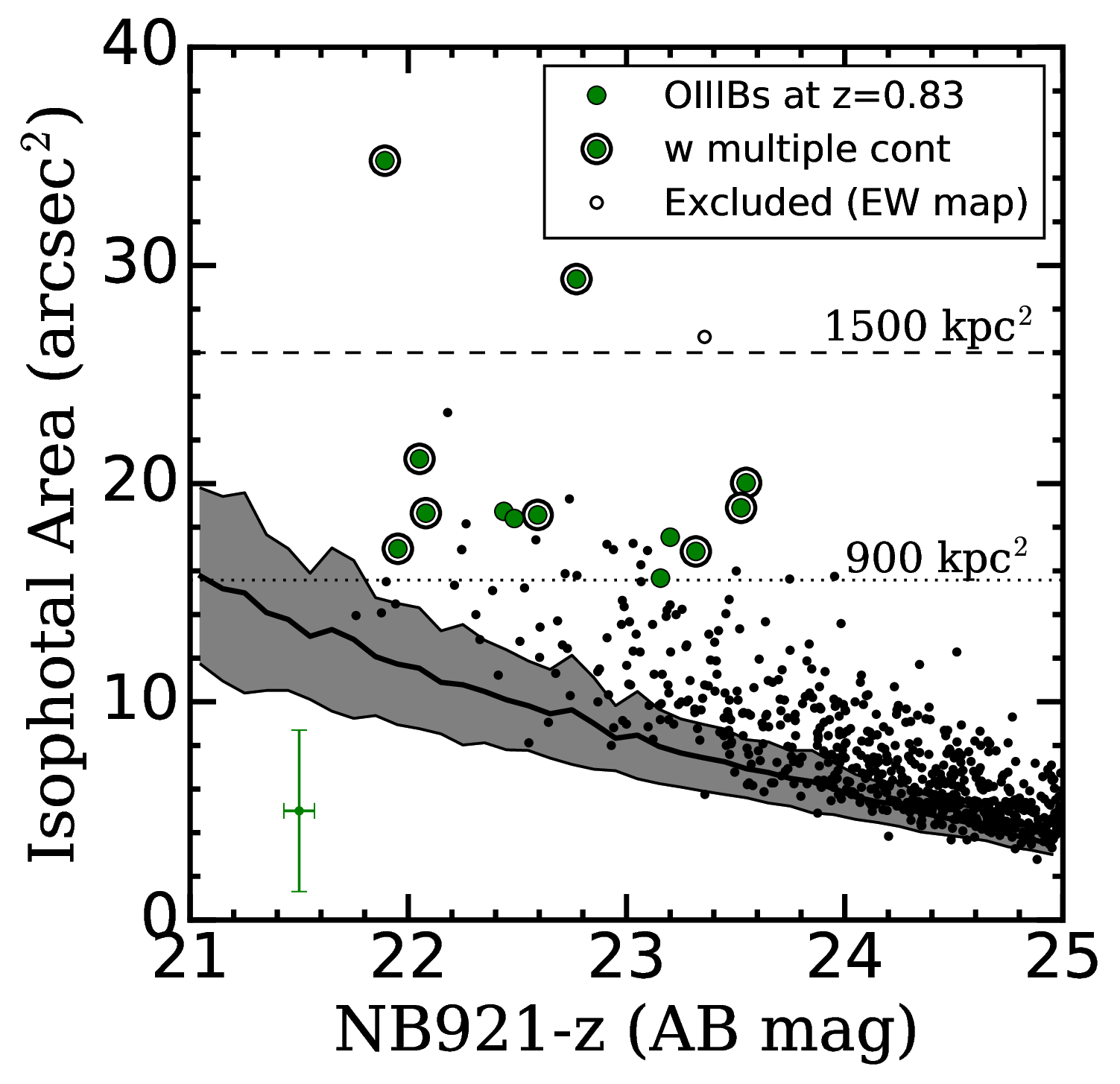} &
		\includegraphics[width=0.3\textwidth]{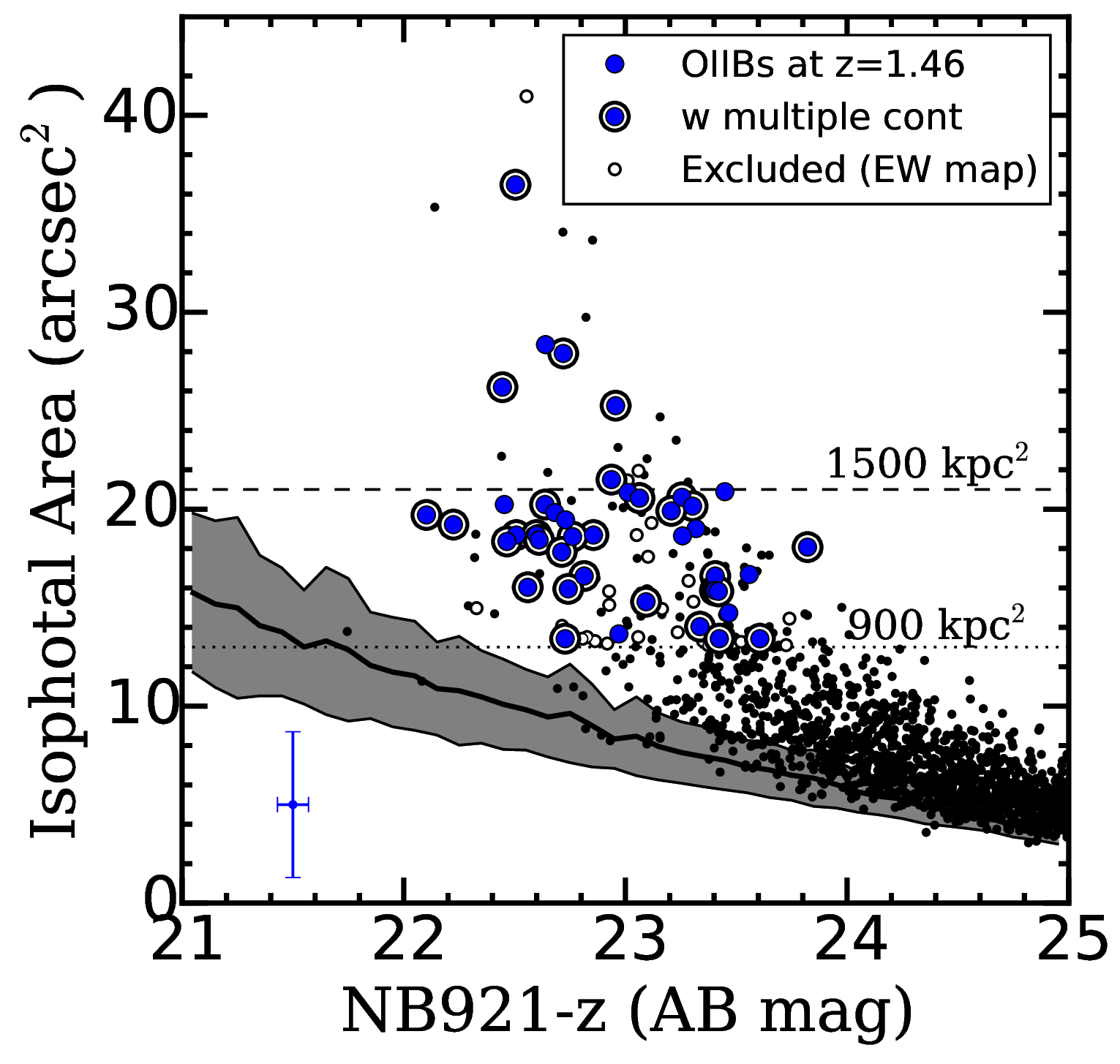}\\
	\end{tabular}
	\caption{Isophotal area-magnitude plot of the {\it \nbnto -\zp} images for 
			\ha\ emitters at $z\sim0.40$ ({\it left}), 
			\oiii\ emitters at $z\sim0.83$ ({\it middle}), 
			and \oii\ emitters at $z\sim1.46$ ({\it right}). 
			Red, green, and blue circles represent the corresponding blobs 
			at respective redshifts. 
			Other symbols are the same as described in Figure \ref{fig_iso_mag_nb816}. 
		}
		\label{fig_iso_mag_nb921}
	\end{figure*}

The isophotal areas of emitters at all redshifts are measured with SExtractor \citep{bertin96} 
down to $1.2\times10^{-18}$ \ergscm\,arcsec$^{-2}$ to match the flux limit 
adopted in Y13. 
Figures \ref{fig_iso_mag_nb816} and \ref{fig_iso_mag_nb921} show 
the magnitudes and isophotal area of all emitters in the emission-line \nbest-\rz\ 
and {\it \nbnto - \zp} images, respectively.\footnote{Note that we do not show the isophotal area plots of 
{\it \nbfot$-$V} and {\it \nbfso$-$V}, because no emitter in these two 
narrowband filters shows the isophotal area larger than 900 kpc$^2$.} 
The magnitudes in the \nbest$-$\rz\ and \nbnto$-$\zp\ images 
indicates fluxes of the continuum-subtracted emission line. 
It is important to note that the object with higher flux (lower magnitude) in the 
\nbest$-$\rz\ or \nbnto$-$\zp\ image is not necessarily more luminous 
in the broadband image, which shows the stellar continuum. 
We separately plot \ha, \oii, and \oiii\ emitters, 
in different panels for clarity. 
The uncertainty of the isophotal area is determined 
by Monte Carlo simulations as in Y13. 
We cut out postage-stamp images of the blobs and place them 
randomly on $\sim1000$ sky regions of the emission-line image 
and re-measure their isophotal areas. 
The errors shown in the figures are 1$\sigma$ of the isophotal area distributions. 

According to the figures, the isophotal area increases 
with the brightness of the emitters, but 
most of the emitters have small isophotal area in the emission-line images. 
A blob candidate is an emitter whose isophotal area in 
the emission-line image is larger than 900 kpc$^2$ or 30-kpc spatial extent. 
We also select the blob candidates with extremely large extension with 
the higher isophotal area criterion of 1500 kpc$^2$. 
Some emitters with isophotal area larger 900 kpc$^2$ 
are excluded during the methods explained below.

\subsubsection{Quantitative approaches to securely select blobs}\label{new_method}

Bright objects and merging systems can mimic the appearance of 
extended emission in the emission-line images. 
If the object is bright enough, it can be falsely classified as a blob candidate. 
On the other hand, small emitters located very close to each other can 
be seen as one extended object in the smoothed emission-line images. 
It is therefore necessary to carefully examine every single blob candidate. 
In Y13, they visually verified the continuum components of 
each blob candidate. However, visual verification is 
time consuming and is potentially affected by personal bias. 
In this paper, we develop the following quantitative methods to 
identify the blobs with genuinely extended emission. 

\begin{itemize}

\item{\it Discrimination of bright PSF-like emitters}

Isophotal area of an object normally depends on its brightness; 
the brighter object tends to show larger isophotal area 
as seen in Figures \ref{fig_iso_mag_nb816} and \ref{fig_iso_mag_nb921}. 
Bright compact emitters may have large isophotal area. 
In order to distinguish the bright PSF-like emitters 
with large isophotal area from the blobs, we create a mock sample of PSFs with 
uniform distribution of emission-line magnitudes ranging from 20 to 25 mag 
and insert them randomly into the original emission-line images. 
We reselect the artificial objects and examine their isophotal area 
and magnitudes using the identical criteria as used for emitters. 
The resulting $1\sigma$ distributions are shown as the shaded area 
in the figures. 
We exclude all emitters falling into the shaded regions from final samples of blobs.

\begin{figure*}
	\centering
	\begin{tabular}{cc}
		\includegraphics[width=0.4\textwidth]{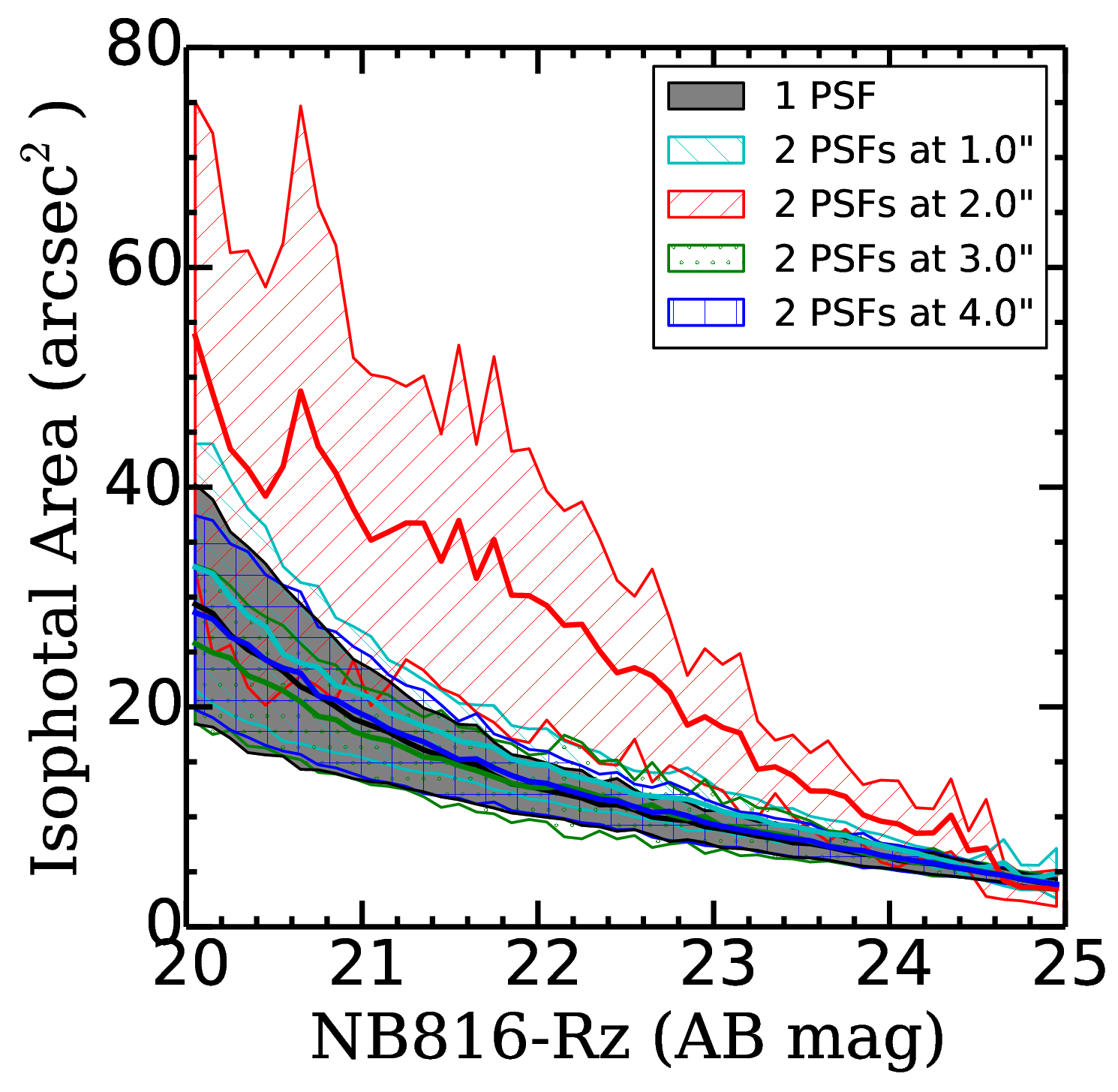} &
		\includegraphics[width=0.4\textwidth]{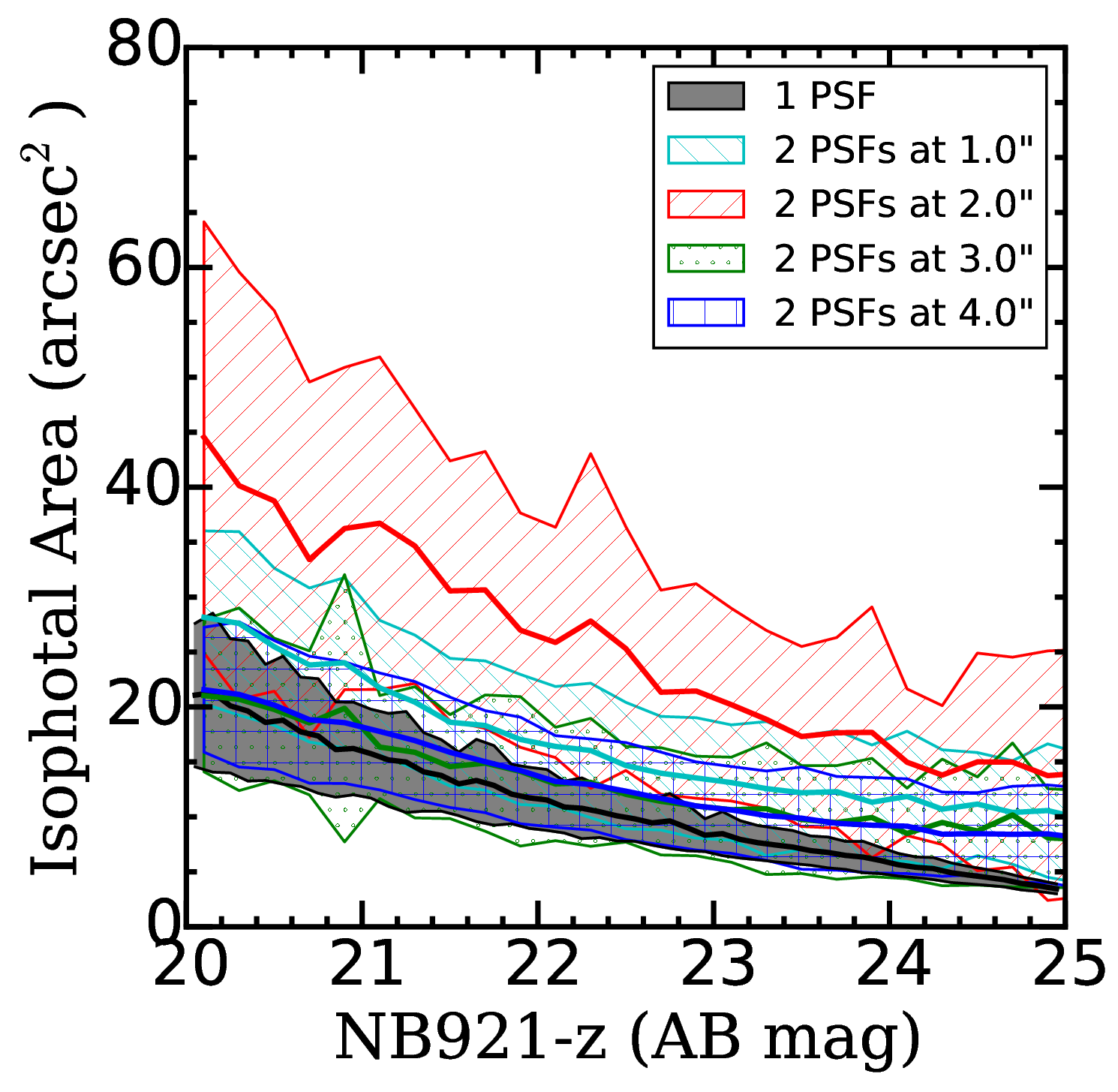}
	\end{tabular}
	\caption{Isophotal area-magnitude plot of 
	two simulated PSFs put close to each other in 
	\nbest-\rz\ ({\it left}) and {\it \nbnto -\zp} ({\it right}) images. 
	The two PSFs are simulated with identical magnitude 
	separated at 1.0\ar, 2.0\ar, 3.0\ar, and 4.0\ar\ distances from each other. 
	The average isophotal areas are shown in thick lines, while the 
	shaded regions represent 1$\sigma$ distributions. 
	Simulation of one PSF is also shown with grey shaded region. 
	}
	\label{fig_sim_psf}
\end{figure*}

Another issue in selecting blob candidates is nearby objects. 
If there is an object located close to the emitters, 
they could appear as one extended object in the emission-line images 
and can be falsely classified as a blob. 
We examine the effect of multiple objects located close to each other 
by simulating two closeby PSFs either the same 
magnitudes separated at various distances or 
with different magnitudes separated at a fixed distance. 
Figure \ref{fig_sim_psf} shows $1\sigma$ standard deviation of 
the isophotal area of the simulated PSFs. 
With the full width at half maximum (FWHM) of $\sim$0.8\ar, 
the isophotal area of the artificial PSFs 
becomes largest at 2.0\ar\ separation and goes down to be approximately 
identical to the isophotal area of one PSF at smaller or larger separations. 
At a distance larger than 2.0\ar\ two PSFs are 
detected as separate sources. 
Their isophotal areas are thus similar to those of one PSF. 
The results are common in all emission-line images 
regardless of the observed wavelengths. 
As a 2.0\ar\ separation of PSFs gives the largest isophotal area, 
we further simulate two PSFs with varying magnitude 
at the fixed 2.0\ar\ separation. 
We found that the isophotal area tends to decrease 
and become comparable to that of one PSF when we increase 
the magnitude difference between two simulated PSFs. 
We apply these results to blob candidates with nearby continuum components. 

After the emitters pass the isophotal criterion, 
we first check the number of continuum components each blob candidate has. 
If it has only one continuum counterpart, we examine if its isophotal area exceeds 
the 1$\sigma$ isophotal-area distribution of the PSF (grey shaded region in the figures). 
If the blob candidate has two continuum components, we further determine 
the separation between those components and check if the isophotal area 
of the candidate is larger than the results of PSF simulations (Figure \ref{fig_sim_psf}). 
The candidates that do not satisfy these PSF criteria are shown as the black dots 
above 900 kpc$^2$ in the figure. 
Finally, if there are more than two components, we will use the next method to 
classify whether the emission is spatially extended or not. 
The blobs with multiple components that pass all criteria are indicated with colored symbols 
enclosed by thick black squares and circles 
in Figures \ref{fig_iso_mag_nb816} and \ref{fig_iso_mag_nb921}, respectively. 

\item{\it Final selection of blobs using equivalent width map}

To further secure the blob sample, 
we develop a new approach to prove the emission-line extension 
beyond the stellar component. 
We construct an equivalent width (EW) map for the individual blob candidate 
obtained after passing the above criteria. 
The observed equivalent width of an emission line is generally defined as 
${\rm EW_{obs}} = F_{\rm line}/f_{\lambda,{\rm cont}}$, where 
$F_{\rm line}$ and $f_{\lambda, {\rm cont}}$ are flux of the emission line 
and flux density of the continuum at the same wavelength, respectively. 
We use the emission-line images listed in Table \ref{tab_emission_image} 
to estimate fluxes of the emission lines 
and the corresponding broadband images to construct the EW map. 
The basic idea of using the EW map is that the EW of the extended 
region is enhanced due to relatively low flux density of the stellar continuum.

\cite{kornei12} studied star-forming galaxies at $0.7<z<1.3$ in the 
Extended Groth Strip (EGS) field and reported the average rest-frame 
\oii\ EW of 45 \AA. 
More than 90\% of their sample shows an \oii\ EW less than 100 \AA\ 
(their Figure 1). 
The star-forming region, where the stellar component is located, 
is unlikely to produce the \oii\ emission with the rest-frame EW as high as 100 \AA. 
Therefore, we adopt the rest-frame \oii\ EW criterion of $>100$ \AA\ 
to define the isophotal area as real extended emission. 
For the \oiii\ emission line, we apply the ratio between 
\oii\ and \oiii\ emission lines of \oiii/\oii$=0.3$, which 
is the typical line ratio of star-forming galaxies at $z\sim1.2$ 
\citep{harikane14} 
and adopt \oiii\ EW larger than $30$ \AA\ as a criterion. 
The \ha\ EW criterion is derived 
from the \oii\ EW by using the relation between SFR and 
the emission lines by \cite{kennicutt98}.  
It is noteworthy that the EW value above criteria 
in the EW map does not directly correspond to the EW of the blob. 
It simply indicates that the pixel with that EW is either the extended emission 
with no/faint stellar component or just noise of the image. 

\begin{figure*}
\centering
\begin{tabular}{cc}
\includegraphics[width=0.4\textwidth]{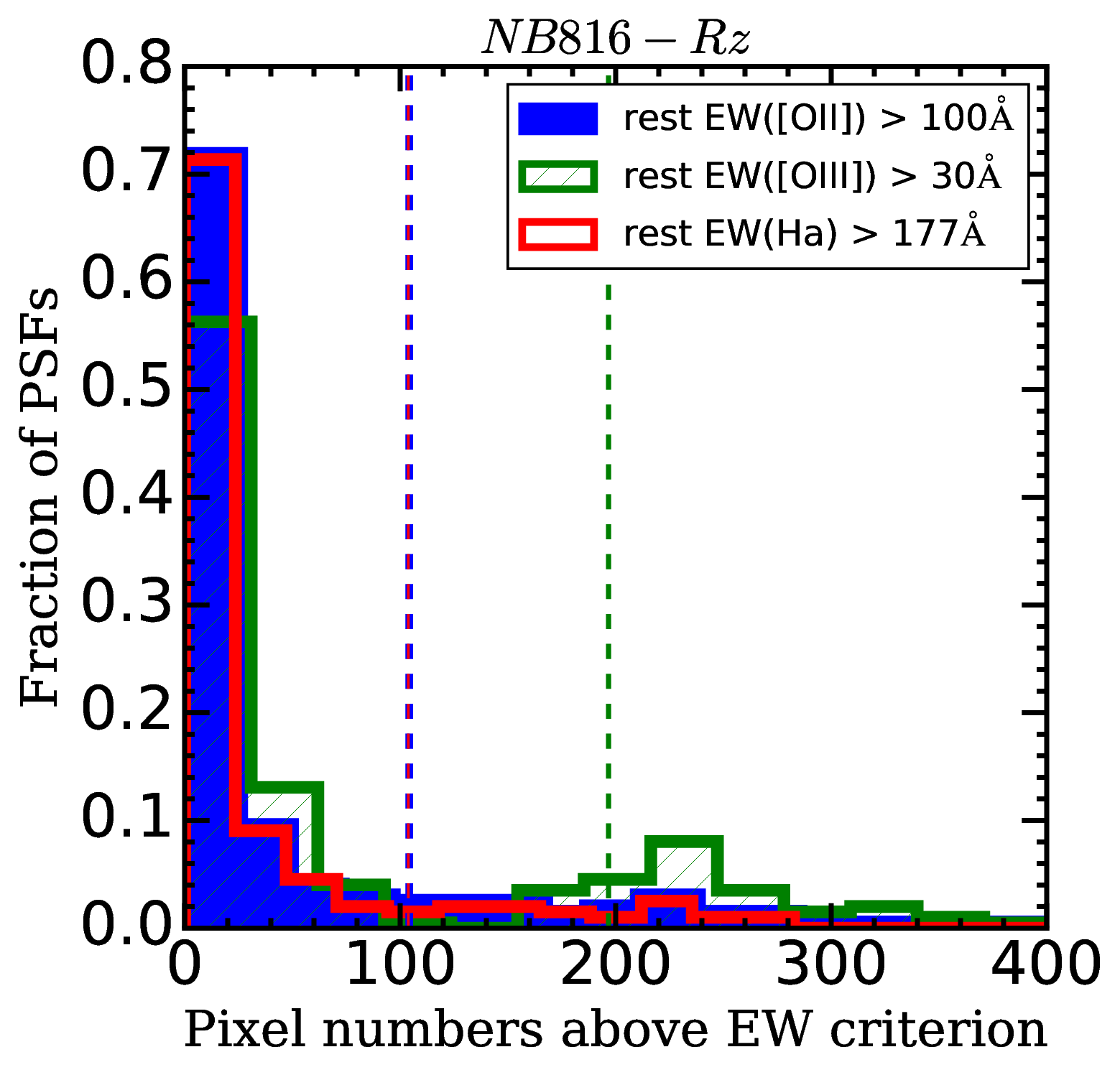} &
\includegraphics[width=0.4\textwidth]{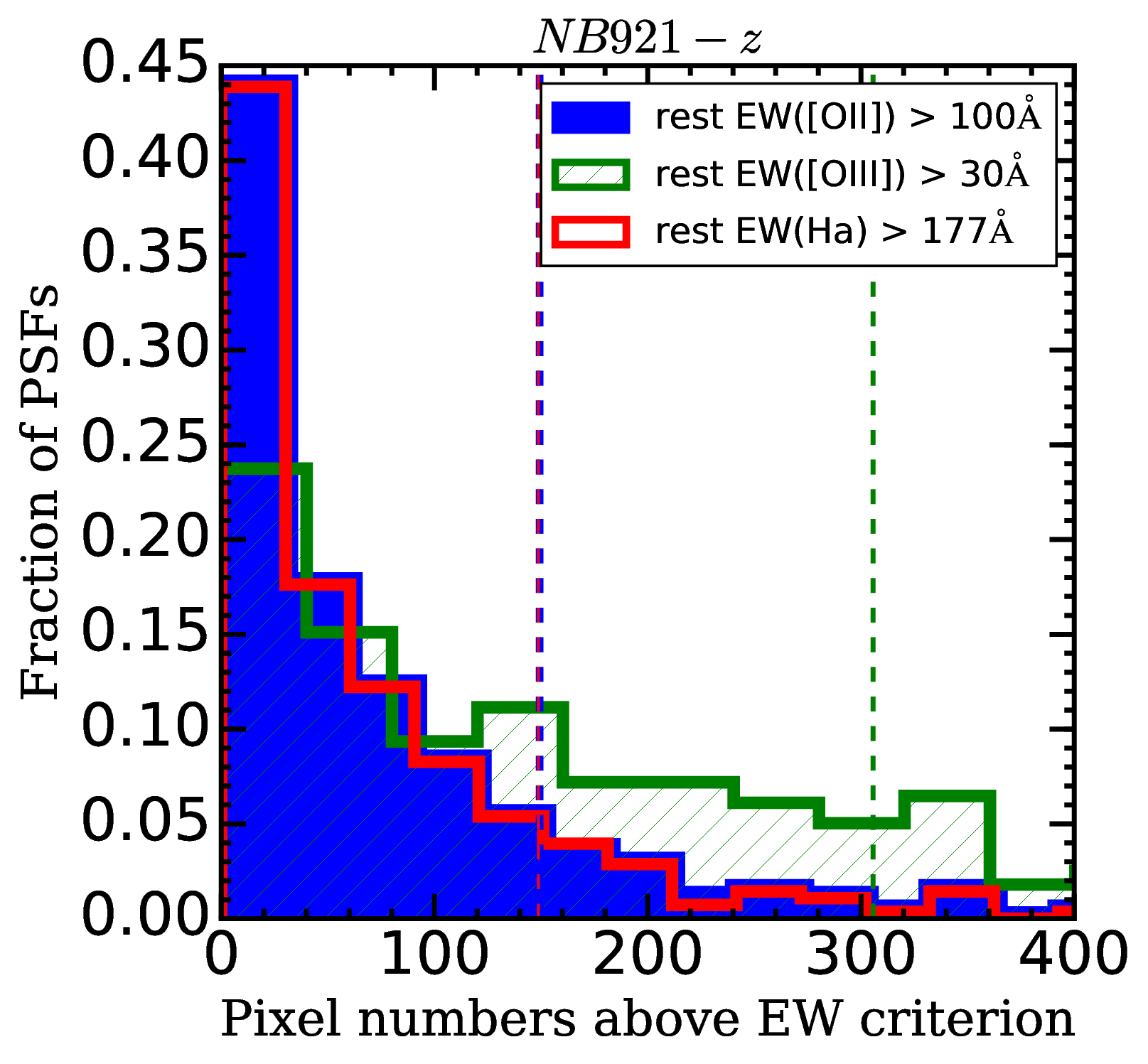} \\
\end{tabular}
\caption{Normalized histograms of the simulated PSFs with the pixels that have the EW 
above the criteria for \oii\ (blue), \oiii\ (green), and \ha\ (red) emission lines 
in the \nbest-\rz\ ({\it left}) and \nbnto-\zp\ images ({\it right}). 
The number of pixels in one PSF is shown on the x axis. 
Dashed lines indicate $1\sigma$ standard deviations that are used as the EW criteria. 
Note that the blue and red histograms are almost identical to each other and are difficult 
to distinguish by eyes. 
}
\label{fig_psf_ewmap}
\end{figure*}

In case of a compact, PSF-like source, the number of pixels with an EW higher than 
the above criteria is ideally supposed to be zero, because of no extended emission line. 
In fact, it is not exactly zero due to the noise around the object. 
We investigate the distribution of the PSFs containing the pixels with high EWs 
in the emission-line image 
in order to determine the minimum number of pixels whose EW is enhanced by the noise in one PSF. 
We create EW maps of the isolated PSFs with the narrowband magnitudes of $20-24$ mag 
and measure the numbers of pixels with EW higher than the criterion for each emission line. 
Normalized histograms of the PSFs with the numbers of pixels having EW above 
criteria are shown in Figure \ref{fig_psf_ewmap}. 
As a point source with no extended feature, the simulated PSFs should not have any pixel with high EW. 
The number of pixels that have the EW larger than the criteria in the PSFs 
can be considered as the possible noise we could obtain when counting the pixels with high EW. 
Almost more than 60\% of the simulated PSFs in the \nbest-\rz\ image 
have less than 40 pixels with the EW higher than the criteria. 
However, a fraction of PSFs contains larger numbers of pixels with the EW above the criteria. 
For the \nbnto-\zp\ image, the simulated PSFs have more pixels with the EW above the criteria, 
but a majority of them still show small numbers of pixels with high EW. 
The blob with a genuinely extended emission line 
should consist of pixels with large EW more than at least $1\sigma$ 
distribution of the PSF noise (thick vertical dashed lines in Figure \ref{fig_psf_ewmap}).

Figure \ref{fig_fake_blob} shows comparison between  \oiib 1 at $z\sim1.2$ by Y13 
and the normal \oii\ emitter with no extension of the emission 
at the same redshift that can be falsely identified as a blob 
if we only use the isophotal area criterion. 
The \oii\ emitter at $z\sim1.2$ in the bottom panel shows the isophotal 
area of 16 arcsec$^2$ or physically 1108 kpc$^2$ in the \nbest-\rz\ image. 
It is larger than the isophotal area criterion we set for \oiib s at $z\sim1.2$ 
($>13$ arcsec$^2$ or 900 kpc$^2$). 
This \oii\ emitter would be selected as an \oii\ blob according to its isophotal area. 
However, as seen in Figure \ref{fig_fake_blob}, the large isophotal area of the emitter is simply 
caused by multiple nearby objects. 
When we smooth the emission-line images with the Gaussian kernel to reduce the fussy noise, 
the nearby objects are blended together and detected as one huge source. 
We are able to get rid of these objects by using the EW map. 
We set the EW per pixel to be larger than 100 \AA\ for \oii\ emission and 
the number of pixels with the EW higher than this criterion needs to 
exceed 104 pixels (Figure \ref{fig_psf_ewmap}). 
As seen in the EW map in the right panel of Figure \ref{fig_fake_blob}, 
the pixels with the rest-frame \oii\ EW higher than 100 \AA\ are only 
the trivial noise at the edge of isophotal area. 
As a result, this object is not satisfied the EW-map criterion. 
In contrast, \oiib 1 in the top panel of the figure shows 
significantly large number of pixels with rest-frame EW higher than 100 \AA. 
With this method, we are able to accurately identify the blobs with spatially extended emission lines.  
The final sample of blobs is shown with colored symbols 
in Figures \ref{fig_iso_mag_nb816} and \ref{fig_iso_mag_nb921}. 
The emitters that are excluded from the final blob sample because 
of the EW-map criterion are indicated with open black circles.

\begin{figure}
	\centering
	\includegraphics[width=0.45\textwidth]{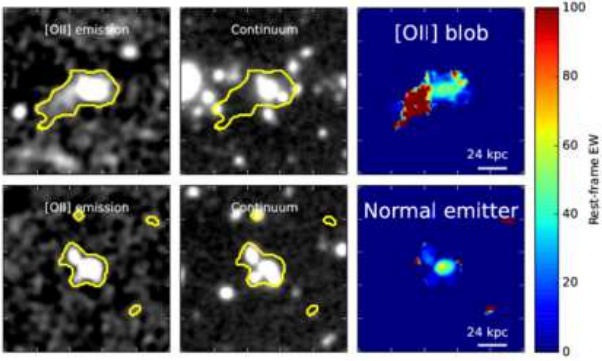}
	\caption{ \oiib 1 at $z\sim1.2$ 
		and the normal \oii\ emitter at the same redshift 
		with multiple nearby objects that make 
		the emitter appear as an extended source in the emission-line image. 
		{\it Left and middle panels:} 
		Emission-line (\nbest-\rz) and continuum (\rz) images. 
		Yellow contours represent the isophotal area of the objects 
		measured in the smoothed emission-line images. 
		{\it Right panel:} The EW maps of the objects. 
		The color bar indicates the rest-frame EW of the emission line 
		in each pixel. 
		The white bar at the bottom right of the figure shows a physical scale of 24 kpc 
		by assuming the redshift at $z\sim1.2$.}
	\label{fig_fake_blob}
\end{figure}

\end{itemize}

\begin{figure}
\centering
\includegraphics[width=0.45\textwidth]{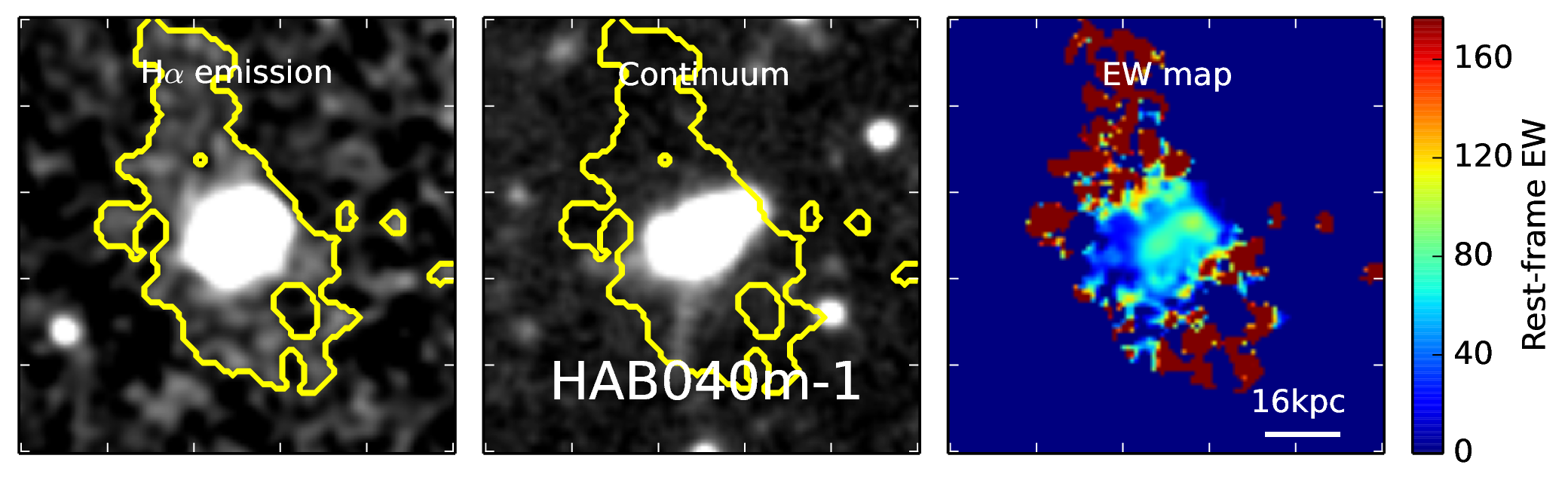}
\caption{From left to right, the emission-line ({\it \nbnto-\zp}), continuum ($z$), 
and EW map images of an \hab\ at $z=0.40$ 
with \ha\ isophotal area larger than 1500 kpc$^2$. 
The ID of the \hab\ is indicated in the middle panel. 
Yellow contours on the left and middle panels show 
the isophotal area of the blob above $1.2\times10^{-18}$ \ergscm\,arcsec$^{-2}$ 
measured in the smoothed emission-line image. 
Note that we do not take the small regions of the contours shown 
in the middle and bottom panels into account when classifying the blobs. 
The color bar on the right panel indicates the rest-frame equivalent width of \ha\ emission.  
The white bar on the right panel represents the physical scale of $\sim16$ kpc. 
In all panels, North is up and East is to the left. }
\label{fig_hab040_final}
\end{figure}

\begin{figure}
\centering
\includegraphics[width=0.45\textwidth]{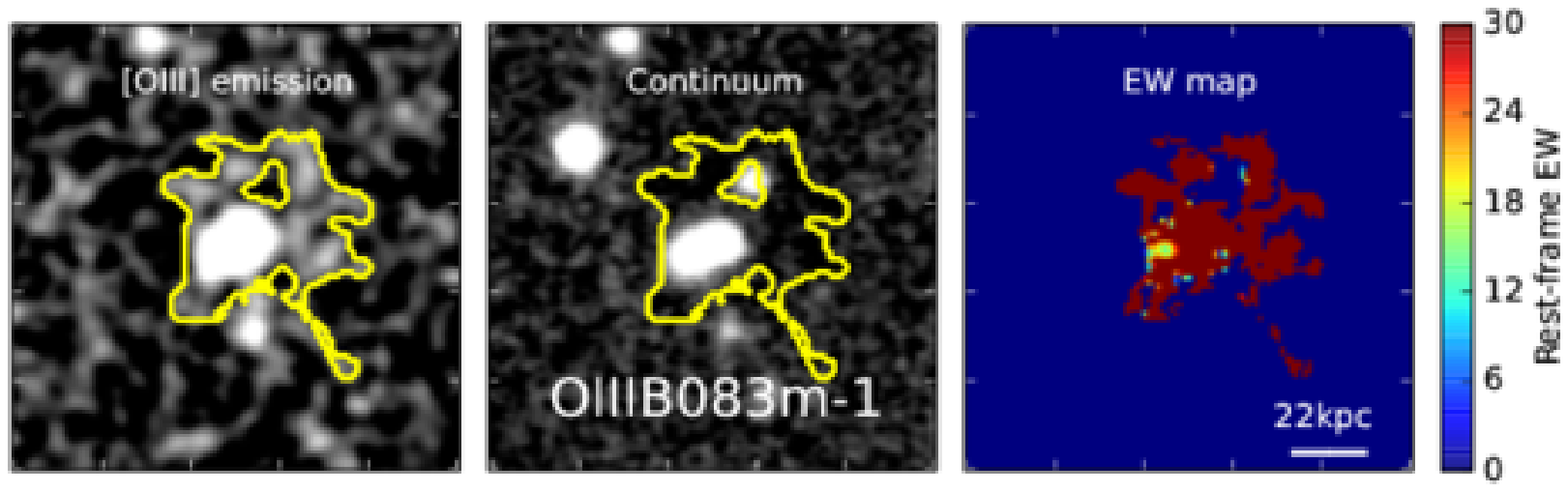}\\
\includegraphics[width=0.45\textwidth]{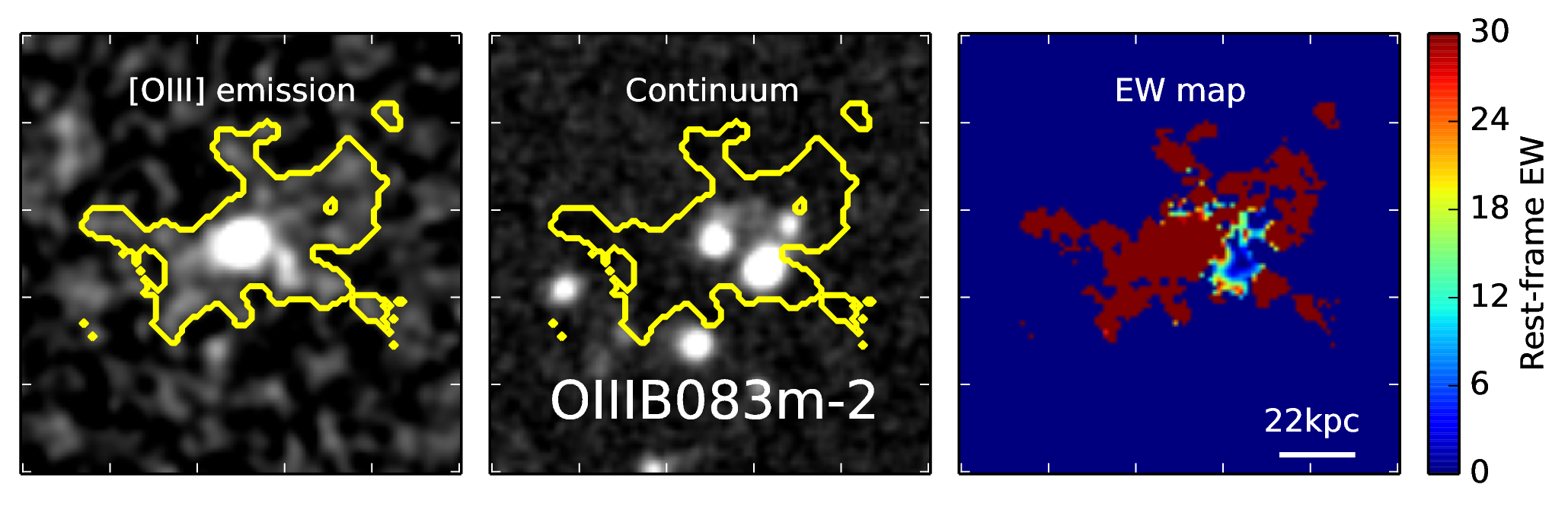}
\caption{Same as Figure \ref{fig_hab040_final} but for \oiiib s at $z=0.83$ 
with isophotal area of \oiii\ emission larger than 1500 kpc$^2$. 
The white bar on the right panel represents the physical scale of $\sim22$ kpc. }
\label{fig_o3b083_final}
\end{figure}

\begin{figure}
\centering
\includegraphics[width=0.45\textwidth]{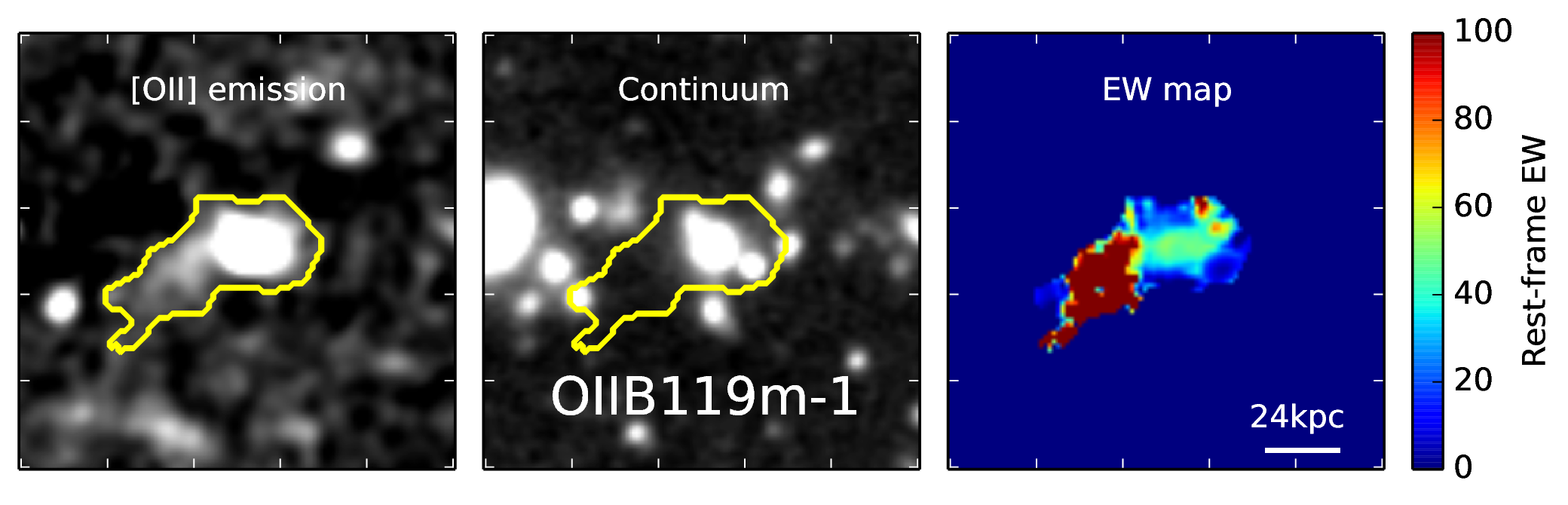}\\
\includegraphics[width=0.45\textwidth]{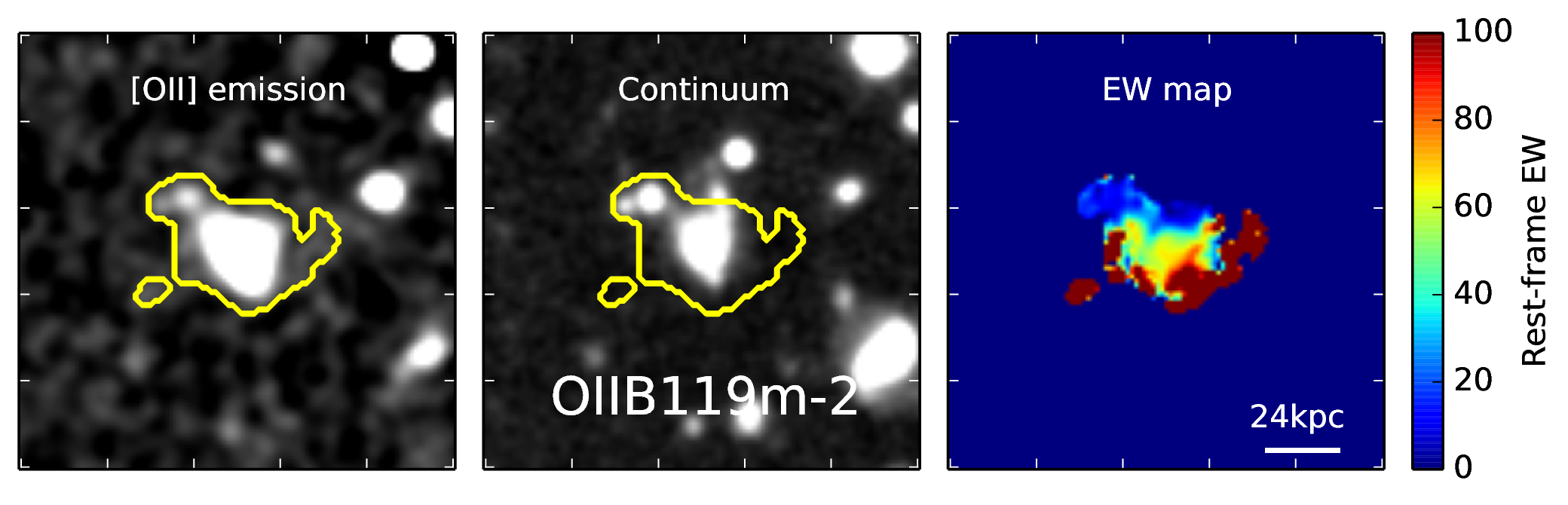}\\
\includegraphics[width=0.45\textwidth]{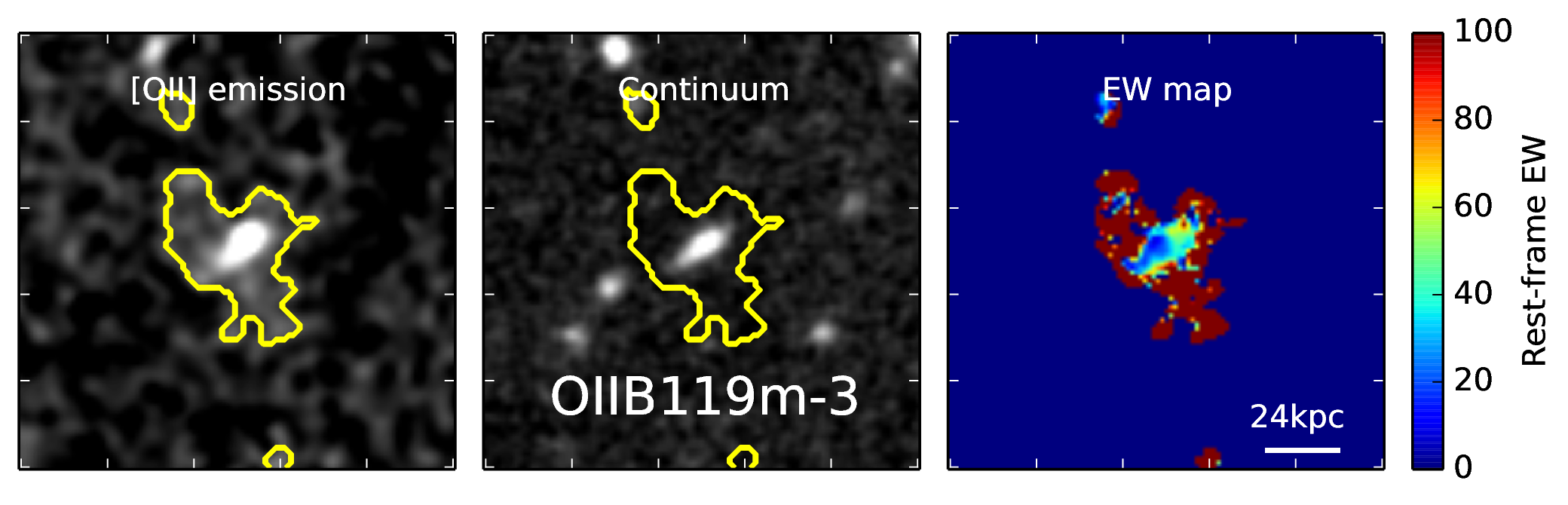}
\caption{Same as Figure \ref{fig_hab040_final} but for \oiib s at $z=1.19$ 
with isophotal area of \oii\ emission larger than 1500 kpc$^2$. 
The white bar on the right panel represents the physical scale of $\sim24$ kpc. 
}
\label{fig_o2b119_final}
\end{figure}

\begin{figure}
\centering
\includegraphics[width=0.45\textwidth]{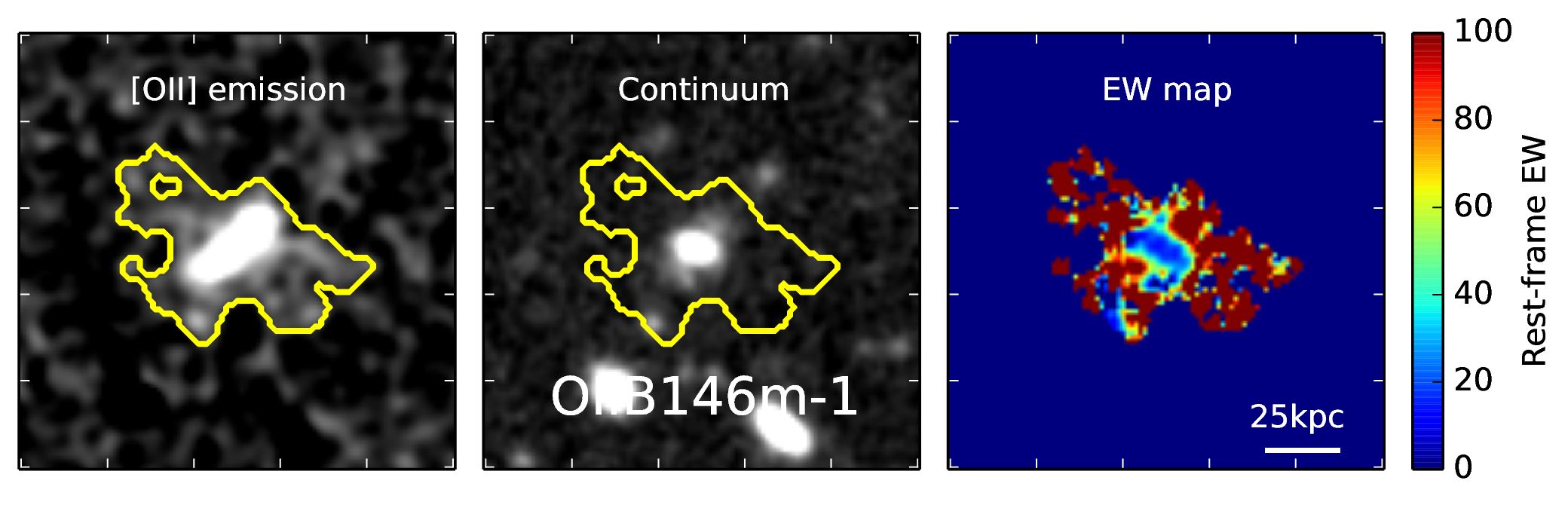}\\
\includegraphics[width=0.45\textwidth]{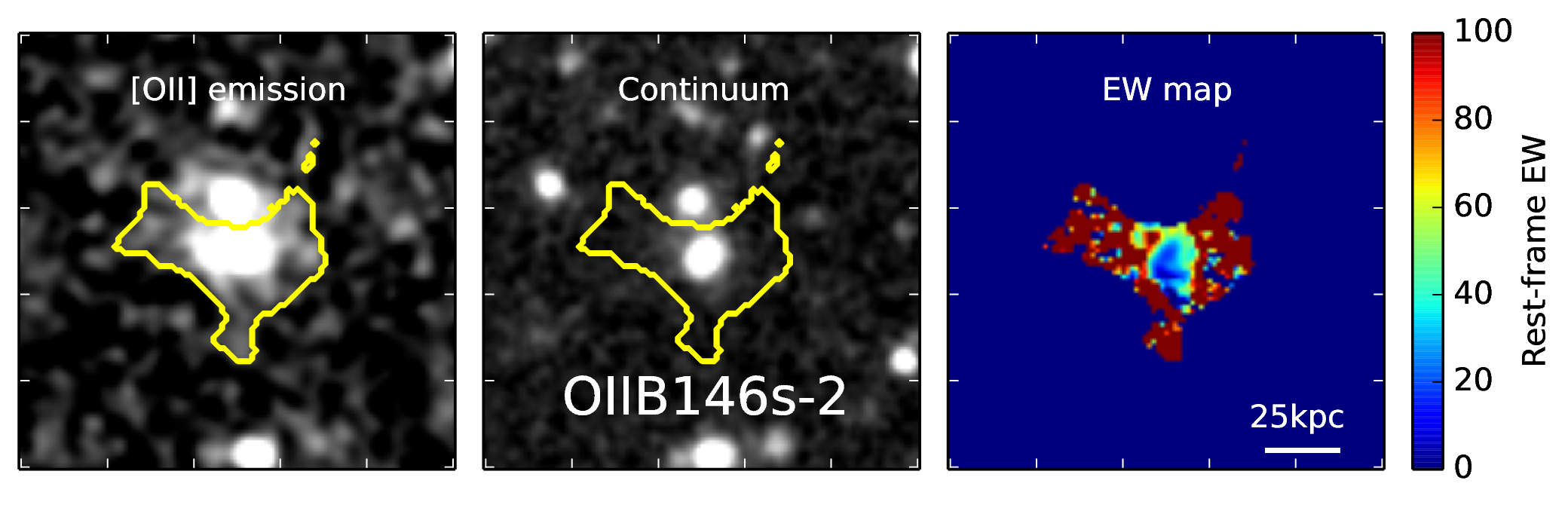}\\
\includegraphics[width=0.45\textwidth]{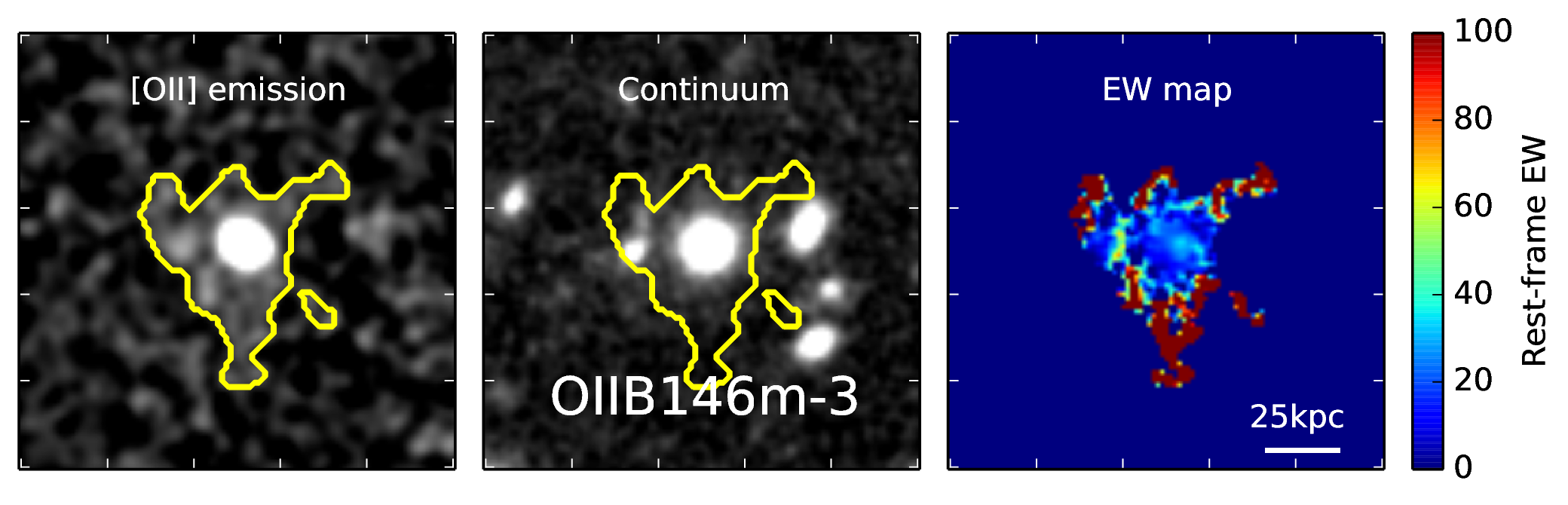}\\

\includegraphics[width=0.45\textwidth]{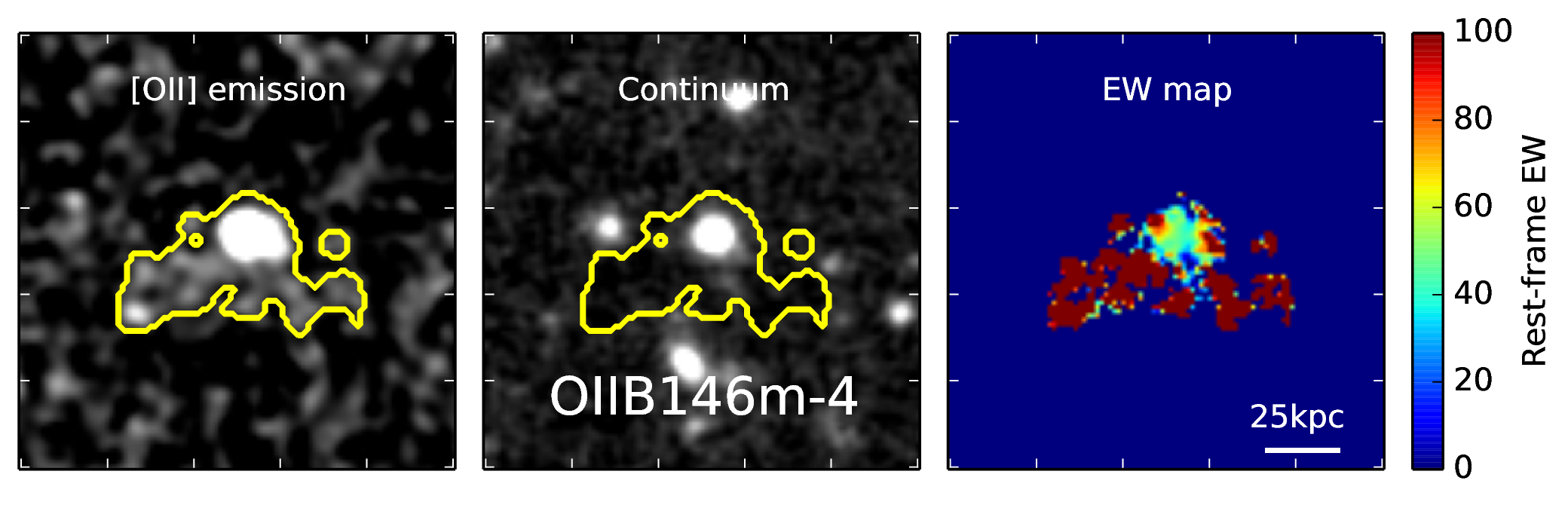}\\
\includegraphics[width=0.45\textwidth]{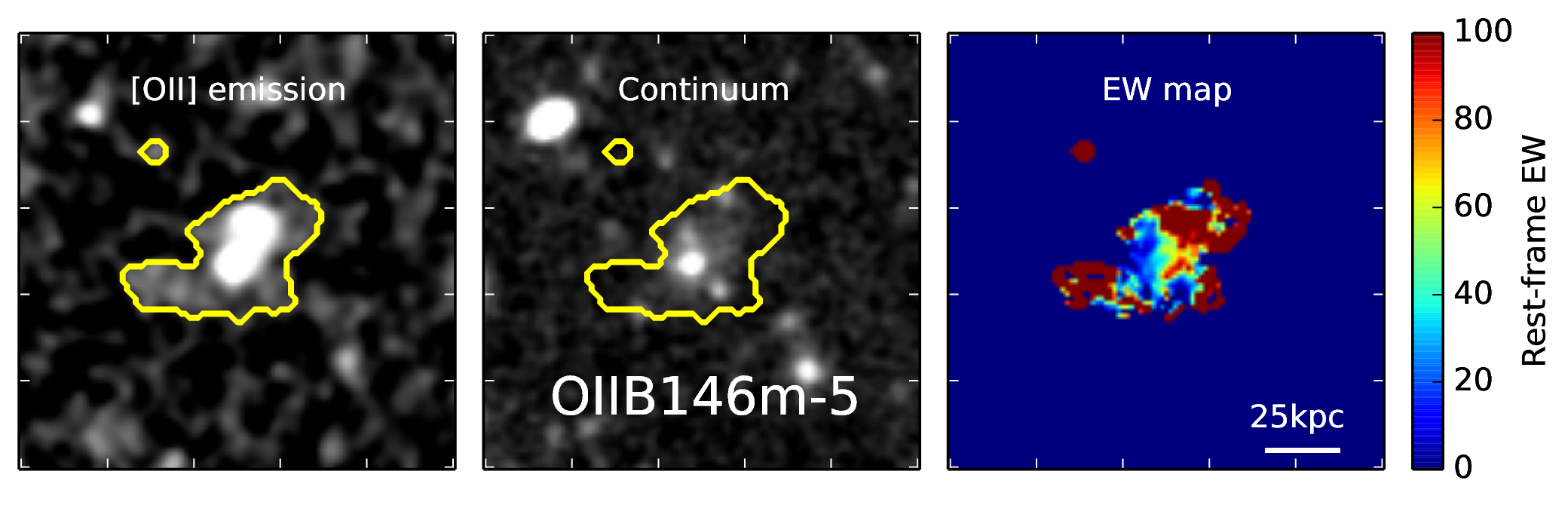}\\
\includegraphics[width=0.45\textwidth]{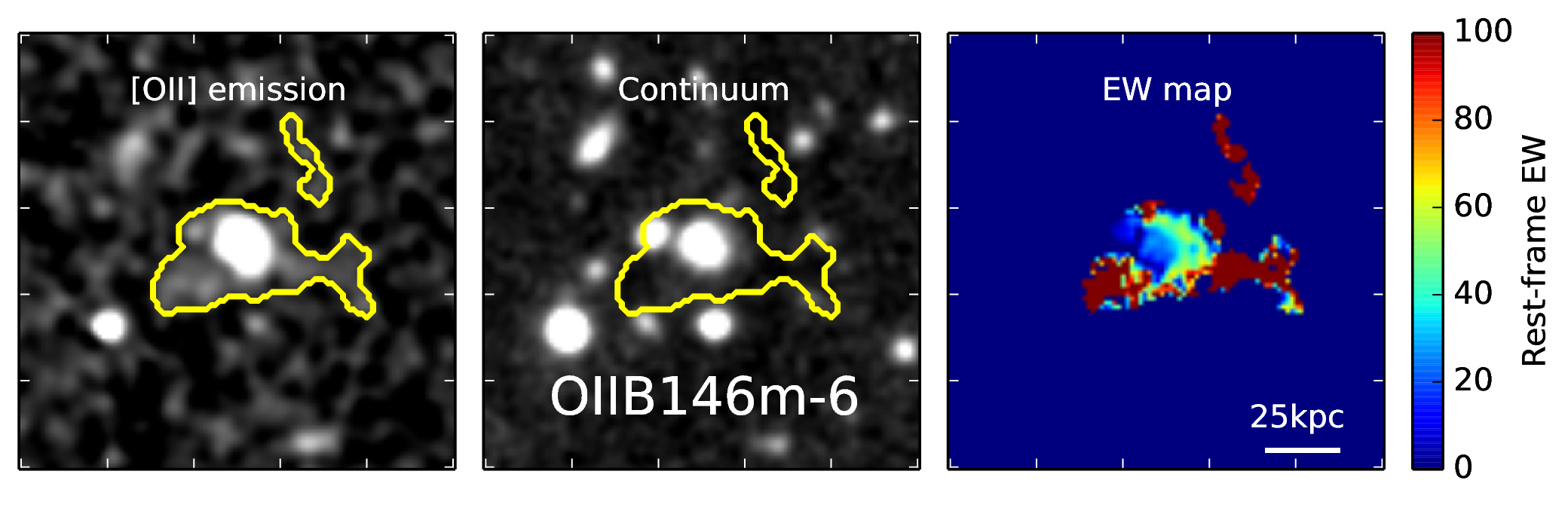}

\caption{Same as Figure \ref{fig_hab040_final} but for \oiib s at $z=1.46$ with isophotal area 
of \oii\ emission larger than 1500 kpc$^2$. 
The white bar on the right panel represents the physical scale of $\sim25$ kpc. 
}
\label{fig_o2b146_final}
\end{figure}

\section{Results}\label{sec:results}

Although we search for the possible blobs with extended \oii, \oiii, and \ha\ emission lines 
from $z\sim 0.14$ to $z\sim1.46$, \ha\ blobs or \hab s at $z=0.40$ are the blob samples 
at the lowest redshift we could acquire. The final samples of 
\oiib s, \oiiib s, and \hab s at $z\sim0.4-1.5$ are summarized in Table \ref{tab_blob_final_sxds}. 
As seen in the table, 
it is obvious that the number of blobs decreases significantly if we consider only the larger isophotal area. 
Coordinates, magnitudes, luminosity, and isophotal area of the emission lines 
of all blob samples are listed in Appendix. 
It is noteworthy that the IDs of blobs are assigned to describe types of blobs
(i.e., \oiib, \oiiib, or \hab), the redshifts, number of stellar components (single or multiple), 
and the number starting from 1 for the largest blob. 
For example, \hab 040m-1 means that this galaxy is the largest \hab\ with multiple stellar components at $z=0.40$. 

\subsection{\hab s at $z=0.40$}\label{subsec:hab040}

At $z=0.40$, there are in total 6 \hab s selected with isophotal area above 900 kpc$^2$, 
but only one object, which we call \hab 040m-1, 
shows the \ha\ emission extended over 1500 kpc$^2$. 
Specifically, the isophotal area of \hab 040m-1 is 1549 kpc$^2$, 
roughly corresponding to 40 kpc across (Figure \ref{fig_hab040_final}). 
This object is spectroscopically confirmed to be at $z=0.407$ \citep{simpson12}. 
The extended \ha\ flux of this blob is $1.18\times10^{-15}$ \ergscm. 
The \ha\ luminosity is $6.94\times10^{41}$ \ergs. 
\hab 040m-1 is also identified as a radio source 
with 1.4-GHz flux density of $\sim120~\mu$Jy 
(Simpson C. in private communication). 
Hence \hab 040m-1 is likely to be powered by an AGN. 
Unfortunately, we do not have the X-ray data covering this object. 
As seen in the middle panel of Figure \ref{fig_hab040_final}, 
\hab 040m-1 seems to show collimated \ha\ emission  
in the direction perpendicular to its continuum component.  
This structure is similar to that of M82, which is 
the prototypical starburst galaxy with large-scale outflow in the local universe 
(see Figure 2 in \citealt{walter02}). 
A follow-up study is necessary to examine the outflow properties and 
the energy source.

\subsection{\oiiib s at $z=0.63$ and $z=0.83$}\label{subsec:o3b}

Four and thirteen \oiii\ blobs or \oiiib s with the isophotal area of 
\oiii\ emission over 900 kpc$^2$ are obtained at $z=0.63$ and $z=0.83$, 
respectively. Among them, only two \oiiib s at $z=0.83$ show 
the extended \oiii\ emission line larger than 1500 kpc$^2$. 
The top panel of Figure \ref{fig_o3b083_final} shows the largest \oiiib\ at $z=0.83$ (\oiiib 083m-1), 
whose \oiii\ emission line covers approximately 2010 kpc$^2$. 
In the bottom panel, the second largest \oiiib, \oiiib 083m-2, shows the extended emission line 
of $\sim1700$ kpc$^2$. As seen in the figure, there are two objects located close to the \oiiib. 
Both objects are not at $z=0.83$ according to their photometric redshifts. 
The EW map in the bottom right panel of Figure \ref{fig_o3b083_final} 
in which the red part indicates the components with 
high \oiii\ EW suggests that these nearby objects are not responsible for the extended \oiii\ emission. 
So the extension is likely to come from the \oiiib.

\subsection{\oiib s at $z=1.19$ and $z=1.46$}\label{subsec:o2b}

At higher redshifts ($z>1$), blobs are selected with the extended \oii\ emission line. 
Y13 firstly discovered 12 \oiib s at $z=1.19$. 
Now we select 11 \oiib s at this redshifts. 
The difference in the \oiib\ numbers between Y13 and our paper is 
mainly due to additional quantitative criteria of 
bright PSFs and EW maps that we apply to ensure the extended feature of \oii\ emission. 
We now have a more secure sample of \oii\ blobs with extended emission lines. 
Figure \ref{fig_o2b119_final} shows three \oiib s at $z=1.19$ with isophotal area 
of the \oii\ emission line larger than 1500 kpc$^2$. 
The \oiib\ with the largest extension of the \oii\ emission line in the top panel of 
Figure \ref{fig_o2b119_final} is \oiib 119m-1, which is spectroscopically identified 
to be an outflowing obscured AGN at $z=1.19$ by Y13 (\oiib 1 in Y13). 
The extended \oii\ emission line is clearly seen in the EW map. 
The middle panel shows the second largest \oiib\ at the same redshift, \oiib 119m-2. 
It is seen in the figure that the area with an EW larger than the criterion is 
not due to the nearby object. 
This object is thus identified as the \oiib. 
The third largest \oiib, \oiib 119m-3, is shown in the bottom panel of Figure \ref{fig_o2b119_final}. 
The stellar component of this \oiib\ appears as an edge-on galaxy, while 
the \oii\ emission line extends in the direction perpendicular to the shape of the continuum, 
similar to the extended \ha\ emission line of \hab 040m-1 in 
section \ref{subsec:hab040} and M82. 

The highest redshift that we search for the blobs in this paper is $z=1.46$. 
Applying the same surface flux density limit at each redshift studied, we 
select 43 objects with the isophotal area of the \oii\ emission line larger than 900 kpc$^2$ 
as \oiib s at $z=1.46$, six of which show even more extended emission over 1500 kpc$^2$ 
(Figure \ref{fig_o2b146_final}). 
The figure is arranged in order of the blob size. 
The \oiib\ with the largest isophotal area is in the top panel of the figure, 
while the smallest one is shown in the bottom panel. 
The largest \oiib\ at $z=1.46$, \oiib 146m-1, is shown in the top panel of Figure \ref{fig_o2b146_final}. 
The extended \oii\ emission line of this \oiib\ is clearly seen. 
The isophotal contour of the second largest \oiib (\oiib 146s-2) shown in the second panel 
from the top of the figure is in the strange shape because it is very close to the other object, 
which is also identified as an \oii\ emitter at the same redshift. 
This is an interesting case in the sense that the \oiib\ may consist of two galaxies 
that are about to merge or in the middle of merging process. 
We can also consider both emitters as one \oiib\ with multiple stellar components. 
It will not affect the total number of \oiib s at $z=1.46$. 
The third \oiib\ \oiib 146m-3 is similar to the first one in the way that it has only one stellar component 
with significant extended \oii\ emission. However, the only difference between these two blobs 
is that \oiib 146m-3 seems to show extended emission in almost all directions, while the \oii\ emission 
in the \oiib 146m-1 looks like bimodal outflow. 
The remaining 3 \oiib s at $z=1.46$ seem to show an asymmetric extension of the \oii\ emission line. 
The shapes of the emission of all \oiib s may suggest that the extended emission can occur 
in different directions depending on the galaxies. 

\begin{deluxetable}{cccc}
\tabletypesize{\footnotesize}
\tablewidth{0pt}
\tablecolumns{4}
\tablecaption{Summary of final sample of blobs in the SXDS field. 
\label{tab_blob_final_sxds}}
\tablewidth{0pt}
\tablehead{
\multicolumn{1}{c}{Redshift} &
\multicolumn{1}{c}{Emission line} &
\multicolumn{2}{c}{Number of blobs\tablenotemark{a}} \\
\cline{3-4}
\multicolumn{1}{c}{} &
\multicolumn{1}{c}{} &
\multicolumn{1}{c}{$>900 {\rm kpc}^2$} &
\multicolumn{1}{c}{$>1500 {\rm kpc}^2$} 
}
\startdata
	  0.40 & \ha$\lambda6563$  &  6 (2/33\%) & 1 (1/100\%)\\\\
	  
	  0.63 & \oiii$\lambda5007$ &  4 (0) &  0 (0)\\
	  0.83 & \oiii$\lambda5007$ & 13 (0) & 2 (0)\\\\
	  
	  1.19 & \oii$\lambda3727$ & 11 (2/18\%) & 3 (2/67\%)\\
	  1.46 & \oii$\lambda3727$ & 43 (4/9\%) & 6 (2/33\%)
\enddata
\tablenotetext{a}{Parentheses show the number of blobs 
with X-ray ($0.5-10$ keV) or radio (1.4GHz) counterpart, which are 
considered to host AGNs, and the percentage of possible AGN fraction.}
\end{deluxetable}

\begin{figure*}
\centering
	\begin{tabular}{cc}
		\includegraphics[width=0.45\textwidth]{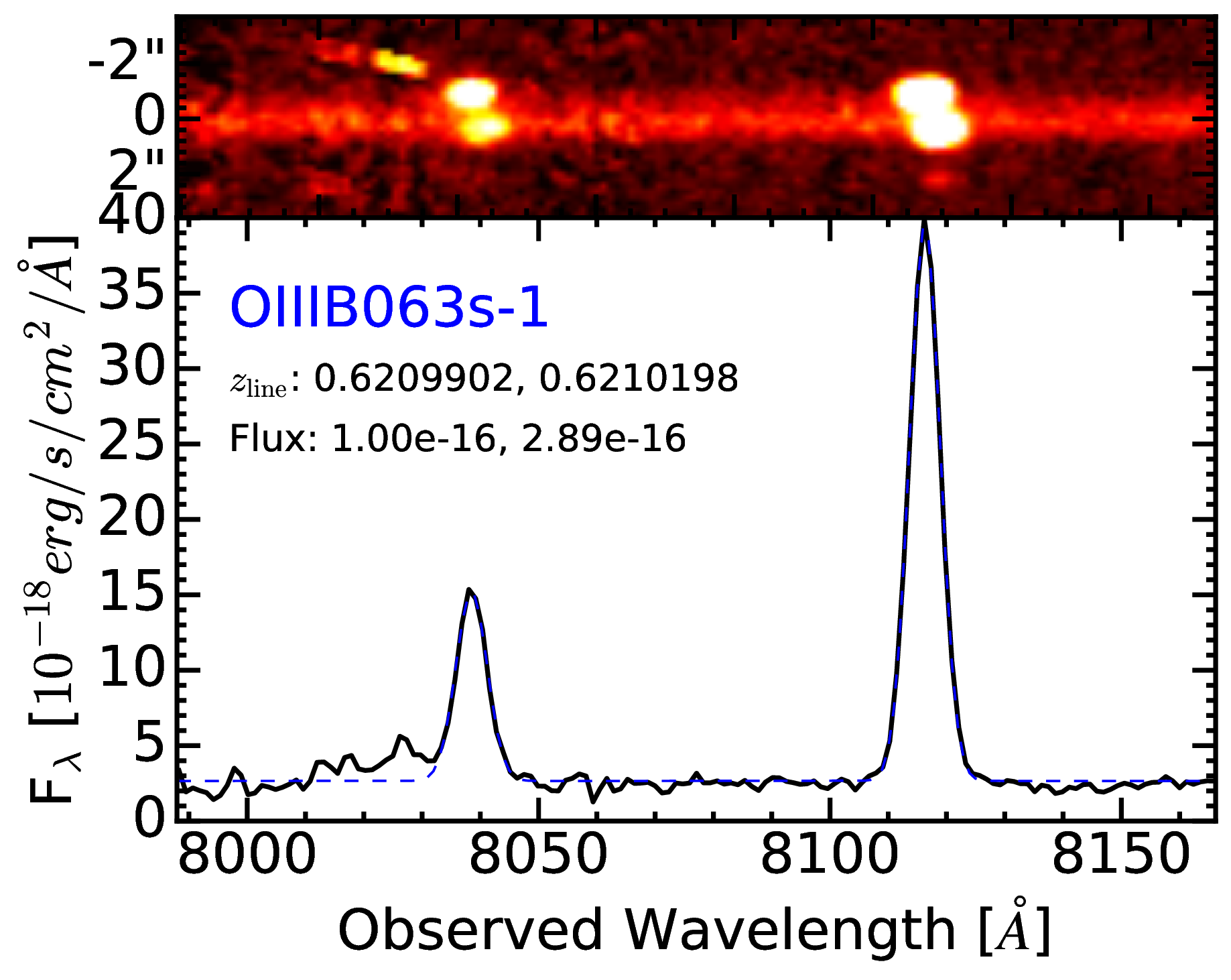} &
		\includegraphics[width=0.45\textwidth]{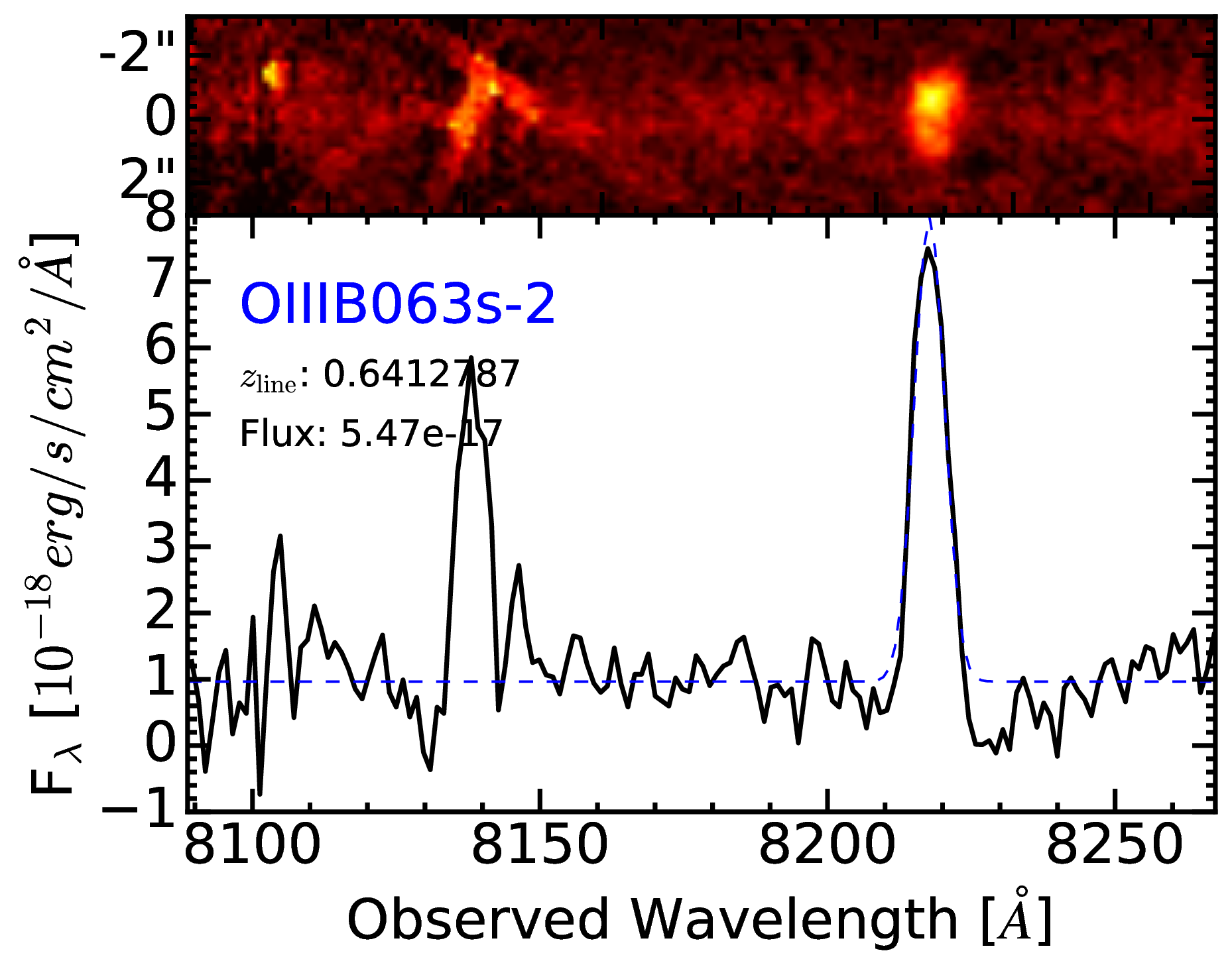} \\
		\includegraphics[width=0.45\textwidth]{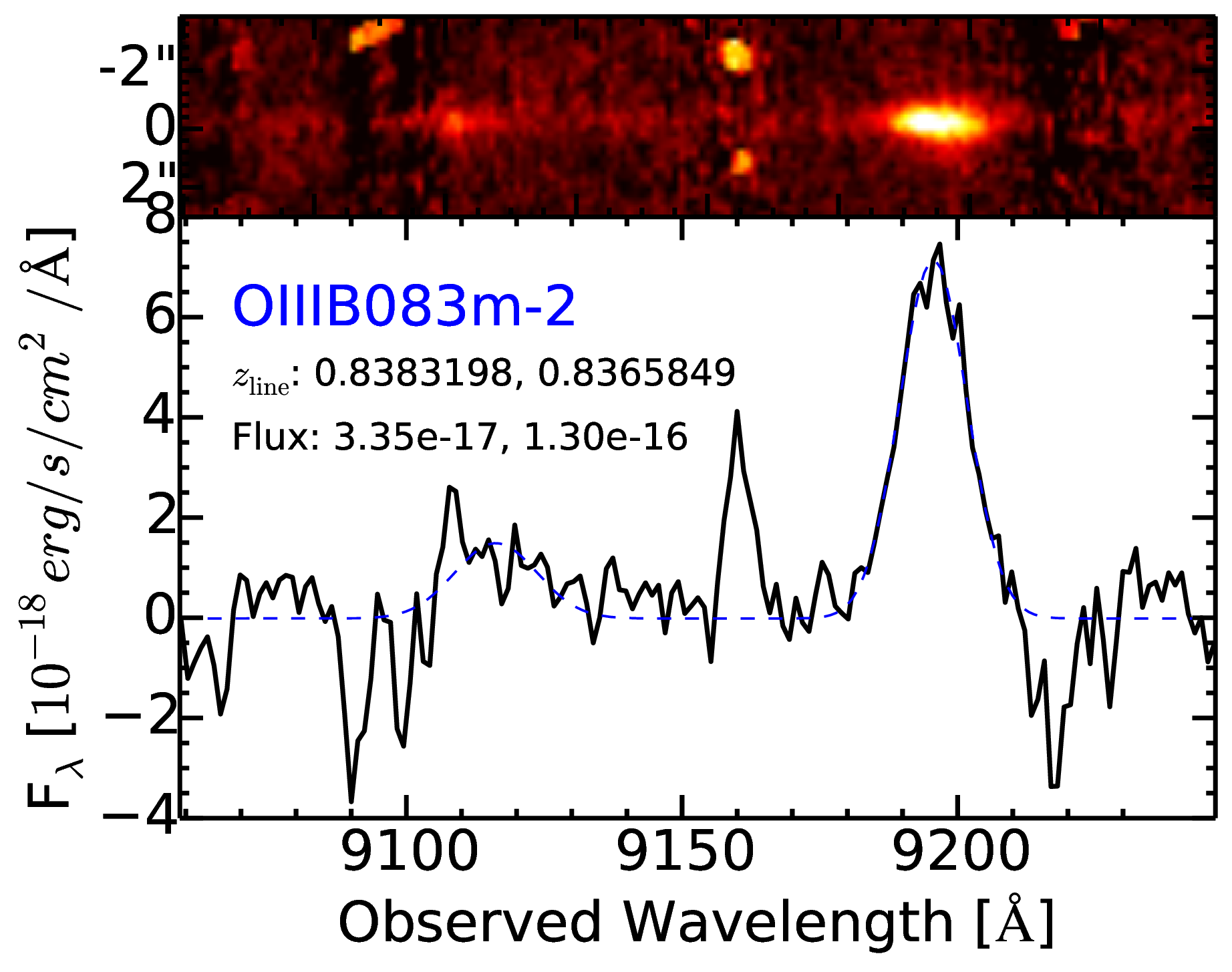} &
		\includegraphics[width=0.45\textwidth]{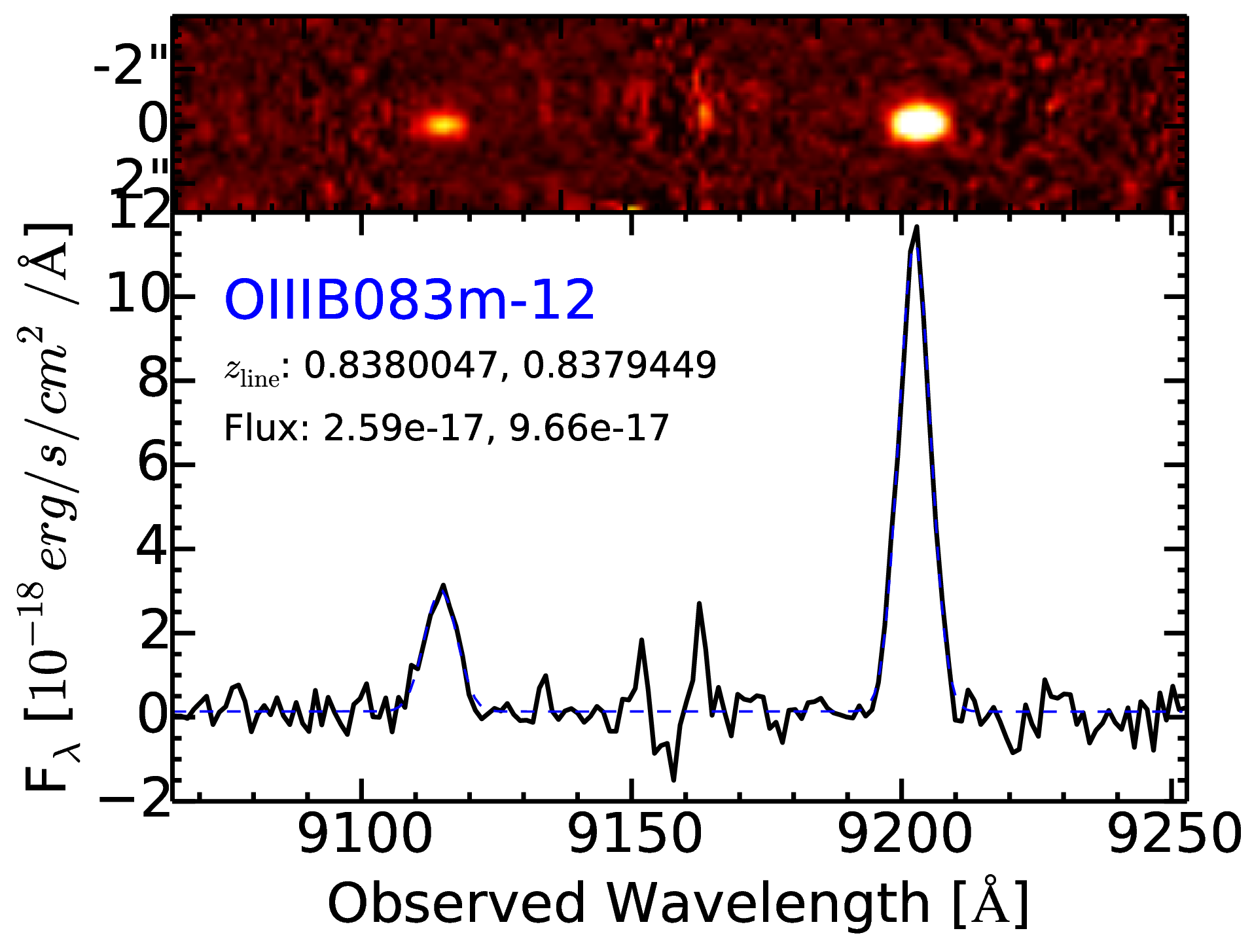} \\
	\end{tabular}
\caption{Subaru/FOCAS Spectra of \oiiib s at $z=0.63$ ({\it top panels}) and at $z=0.83$ ({\it bottom panels}) 
in the observed frame. Each panel shows the \oiiib\ spectrum 
covering \oiii$\lambda\lambda4959, 5007$ emission lines in black solid line and 
the best-fit Gaussian profile in blue dashed line. 
The corresponding two-dimensional spectrum is also shown at the top of the panel. 
Redshifts and observed fluxes of the emission lines are indicated. 
}
\label{fig_spec}
\end{figure*}

\subsection{Spectroscopic Confirmation}\label{subsec:specz}

We perform the spectroscopic follow-up observations for 4 \oiiib s 
at $z=0.63$ and $z=0.83$ (Section \ref{subsec:specdata}). 
Due to the limited observing time, we prioritize our follow-up observations to 
the blobs with the largest extension at each epoch and 
simultaneously maximize the number of blobs in one MOS mask. 
As a result, we obtain the spectra of two largest \oiiib s at $=0.63$ (\oiiib 063s-1 and \oiiib 063s-2), 
the second largest \oiiib\ (\oiiib 083m-2), and one small \oiiib (\oiiib 083m-12) at $z=0.83$. 
Their \oiii\ spectra are shown in Figure \ref{fig_spec}. 
We carefully investigate the possibilities of \lya, \oii, and \ha\ emitters at other redshifts. 
At $z=0.63$, the \oiiib s should exhibit an emission line at $\sim8100$ \AA\ falling into the \nbest\ narrowband filter. 
If it is \oiii$\lambda5007$ emission, we should observe the doublet with 
an \oiii$\lambda4959$ emission at the shorter wavelength. 
For the largest blob at $z=0.63$ (\oiiib 063s-1), we detect another emission line 
at $\sim8045$ \AA, which perfectly corresponds to \oiii$\lambda4959$ emission. 
Furthermore, we also detect the \oii\ emission line at $\sim6040$ \AA. 
Therefore, \oiiib 063s-1 is spectroscopically confirmed to be at $z=0.621$. 
For \oiiib 063s-2, we detect an emission line at $\sim8200$ \AA. 
It is confirmed to be \oiii$\lambda5007$ at $z=0.641$, 
because we detect \oii$\lambda3727$ line at $\sim6120$ \AA.\footnote{
Note that the resolution of the spectrum is insufficient to distinguish the \oii\ doublet.} 
Similarly, we detect \oiii$\lambda4959, 5007$ doublet at $\sim9120$ \AA\ and $\sim9200$ \AA\ 
and the \oii$\lambda3727$ emission at $\sim6850$ \AA for two \oiiib s at $z=0.83$ (\oiiib 083m-2 and \oiiib 083m-12) 
shown in the bottom panels of Figure \ref{fig_spec}. 
They are both at $z=0.838$. 

Two-dimensional spectra are also displayed at the top of each panel in Figure \ref{fig_spec}. 
The \oiii\ emission lines of the \oiiib s are significantly more extended than the stellar continuum, 
especially \oiiib 063s-1. 
Two components of emission line are obviously seen. 
Both components are roughly at the same redshift of $z=0.61$, 
suggesting that \oiiib 063s-1 is probably a merging system, even though 
only one stellar component has been detected in the broadband continuum image. 
Further analysis and discussion on the spatial profiles of emission lines and 
the outflow signature will be in the upcoming paper (Yuma et al. in prep.).

\section{Discussion}\label{sec:discuss} 
\subsection{Primary Check for AGN Fraction}\label{subsec:agn}

An AGN is one of the most energetic objects in the universe. 
It is highly plausible that the \oiib s, \oiiib s, and \hab s at $z=0.4-1.5$ 
we obtain in the previous section 
are partially or entirely powered by the AGN. 
In fact, Y13 found that one of their 12 \oiib s at $z\sim1.2$, 
\oiib 1, is a radio-quiet obscured type-2 AGN. 
In this section, we primarily investigate whether or not 
the spatially extended emission line of the blobs is due to an AGN by cross-checking 
our blobs with X-ray and radio source catalogs. 

The SXDS X-ray data in the $0.2-10$ keV band were taken with 
the European Photon Imaging Camera 
\citep[EPIC; ][]{struder01, turner01} on board {\it XMM-Newton} \citep{jansen01}. 
\cite{ueda08} created the X-ray source catalog from these data down to 
sensitivity limits of $6\times10^{-16}$, $8\times10^{-16}$, $3\times10^{-15}$, 
and $5\times10^{-15}$ \ergscm\ 
in the $0.5-2$, $0.5-4.5$, $2-10$, and $4.5-10$ keV bands, respectively. 
These flux limits roughly correspond to the X-ray luminosities of $10^{41}-10^{42}$ 
\ergs\ at $z=0.40$ and $10^{43}$ \ergs\ at $z=1.46$, which are one order of magnitude 
brighter than the faint end of X-ray luminosity functions (XLFs) of AGNs at similar redshifts 
\citep[e.g., ][]{aird15, ranalli16, fotopoulou16}. 
We identify the X-ray sources that are within 3\ar\ radius of the blobs at $z=0.4-1.5$ 
as the X-ray counterparts. The reason why we use 3\ar\ radius 
is that it is a typical half width at half maximum of the PSFs 
in {\it XMM-Newton} images \citep{struder01}. 
However, we sometimes use larger radius to identify the X-ray counterpart, 
if the position uncertainty of the X-ray source is higher than 3\ar.

The VLA/1.4 GHz radio catalog is provided by \cite{simpson06}. 
The flux density limit is 100 $\mu$Jy. 
Because the typical beam size in the radio image is $5\times4$ arcsec$^2$, we 
crossmatch the radio sources with 2.5\ar\ radius. 

The numbers of blobs that have X-ray and/or radio counterparts are listed in the 
parentheses in Table \ref{tab_blob_final_sxds}. 
We consider the blob as an AGN host if it has either an X-ray or radio counterpart. 
According to Table \ref{tab_blob_final_sxds}, a fraction of blobs which possibly host 
an AGN increases with the isophotal area of the extended emission line. 
This trend is seen in all types of emission lines,  
probably because an AGN is able to provide energy high enough 
to drive the gas out of the galaxy. 
AGNs typically show \oiii/\oii\ ratios larger than 1.0, which is 
higher than the star-forming galaxies with no AGN activity \citep{lamareille10}. 
Thus \oiiib s are expected to host AGNs in a larger fraction than those of other types of blobs. 
However, no X-ray or radio counterpart is found for \oiiib s at $z=0.63-0.83$. 
\oiii\ emission is a high excitation emission line. 
One possibility is that the \oiiib s are too faint to be detected in the 
X-ray catalog we use. As mentioned above that the luminosity limits 
of the catalog are significantly brighter than the faint end of the XLFs, 
many AGNs at the similar redshifts show X-ray luminosity lower than the limits of our catalog. 
Another possibility might be the photoionized light echo from the AGN phase 
that still travels across the galaxy after the black hole at the center 
stops accreting \citep[e.g., ][]{schawinski15}. 
In this case, the extended \oiii\ emission is a leftover from the past AGN activity, 
while the AGN already stops emitting the X-ray after the AGN phase. 
Besides an AGN, there are another possible factors responsible for 
high \oiii/\oii\ ratio such as a low metallicity \citep{overzier09}, 
a high ionization parameter \citep{kewley13}, 
and the density bounded \hii\ region \citep{nakajima14}. 
It is noteworthy that we cannot completely identify all AGNs with only X-ray 
and radio imaging data. 
There are probably \oiib s, \oiiib s, or \hab s hosting X-ray-faint, radio-quiet, 
and/or heavily obscured AGNs. 
Therefore, the AGN percentages listed in Table \ref{tab_blob_final_sxds} 
can be considered as a lower limit of AGN fraction on blobs.

\subsection{Stellar properties}\label{subsec:mass}

Photometry of the selected blobs is obtained 
in 10 bands: {\it BVRizJHK}, IRAC ch1 (3.6\micron) and ch2 (4.5\micron) bands. 
For the photometry in optical and near-infrared wavelengths, 
we perform source detection by using the {\it K}-band image 
and measured the magnitudes at 2.0\ar\ diameter aperture 
at the identical positions in all other images using the dual mode of SExtractor. 
The total magnitudes of the objects are obtained by 
scaling the aperture magnitude in {\it K} band to match {\tt MAG\_AUTO} from SExtractor. 
The photometry in mid-infrared wavelengths (i.e., IRAC 3.6 and 4.5 \micron) 
is determined from the aperture magnitude at 2.4\ar\ diameter, 
which provides the highest signal-to-noise ratio. 
The total magnitude in IRAC is obtained by applying the aperture correction factor that 
is determined by generating artificial objects with given magnitudes on the IRAC images 
and measuring their aperture magnitudes at 2.4\ar\ diameter. 
To avoid the effect of strong nebular emission on the SED fitting, 
we exclude the photometry in \ip\ or \zp band from the fit depending on the narrowband filters 
used to originally select the blobs (i.e., \nbest\ or \nbnto, respectively). 
It is noteworthy that excluding the photometry in either \ip\ or \zp\ band 
does not affect the resulting stellar masses of blobs. 
We make a comparison of the stellar masses with those 
derived by using the photometry in all broadband filters and those using the photometry 
from which the contribution of strong emission lines (i.e., \oii, \oiii, and \ha) is excluded. 
The resulting stellar masses are comparable within 0.3 dex and 0.2 dex, respectively, 
with no systematic offset.

The stellar populations of blobs are examined by fitting the observed photometry 
(spectral energy distributions; SEDs) with the stellar population synthesis models 
constructed with \cite{bc03} code. 
We assume the \cite{salpeter55} initial mass function with standard lower 
and upper mass cutoffs of 0.1 \Msun\ and 100 \Msun, respectively, 
dust attenuation by \cite{calzetti00}, and solar metallicity. 
We adopt the spectroscopic redshifts when available, otherwise 
we fix the redshifts by assuming that the emission line of each type of blobs 
falls into the center of the narrowband filter. 
We create the model SEDs with constant star formation history and perform 
the fitting by using {\tt SEDfit} code by \cite{sawicki12}. 

\begin{figure}
	\centering
	\includegraphics[width=0.45\textwidth]{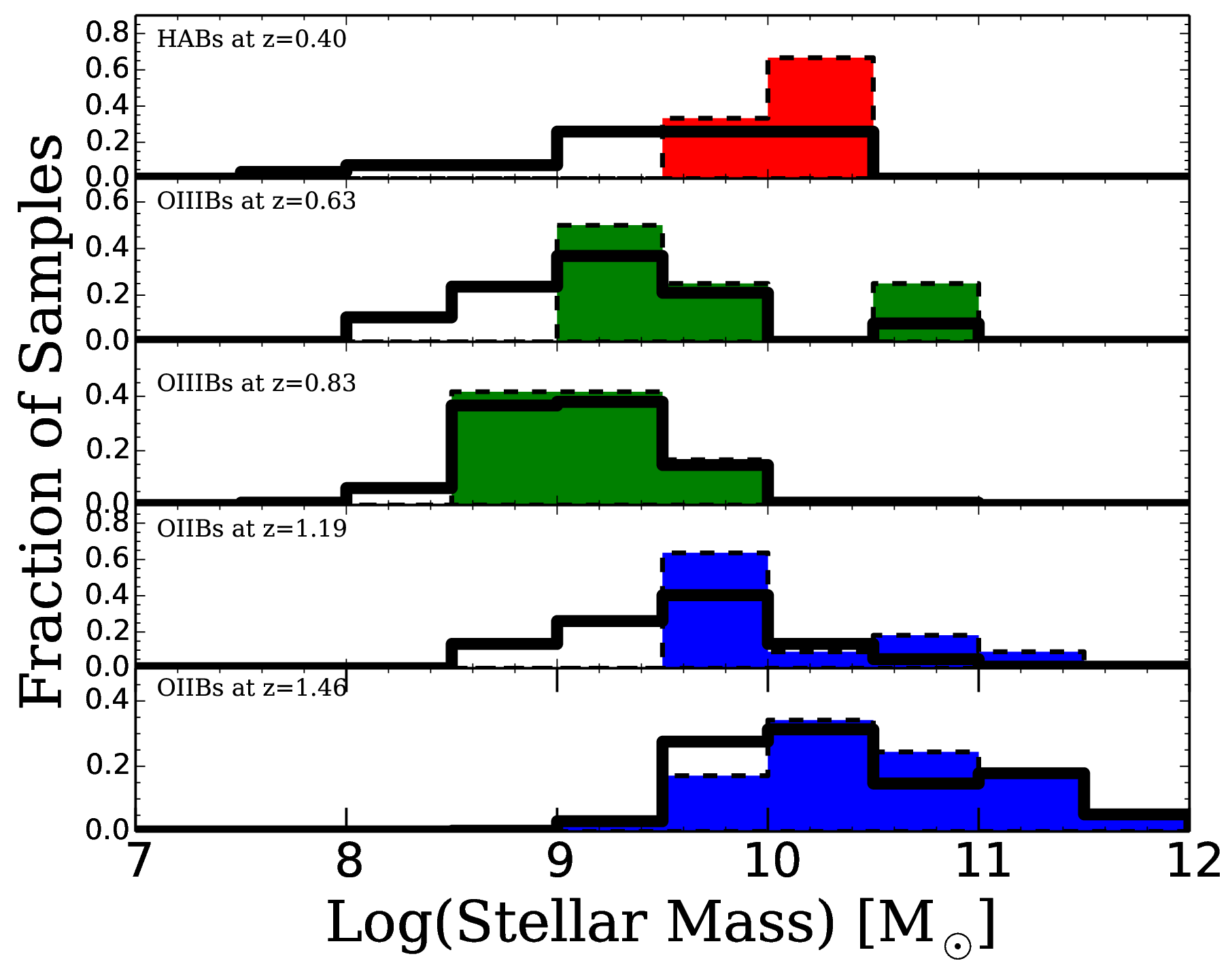}
	\caption{Normalized histograms of stellar masses for blobs and emitters at $z=0.40-1.46$. 
	Blobs are shown with colored histograms, whereas open ones are for the emitters at the corresponding redshifts.
	The histograms are normalize in the way that they show the fraction of blobs/emitters in each mass bin. 
	Types and redshifts of blobs/emitters are indicated in each panel. 
	}
	\label{fig_histogram_mass}
\end{figure}

\begin{figure}
	\centering
	\includegraphics[width=0.45\textwidth]{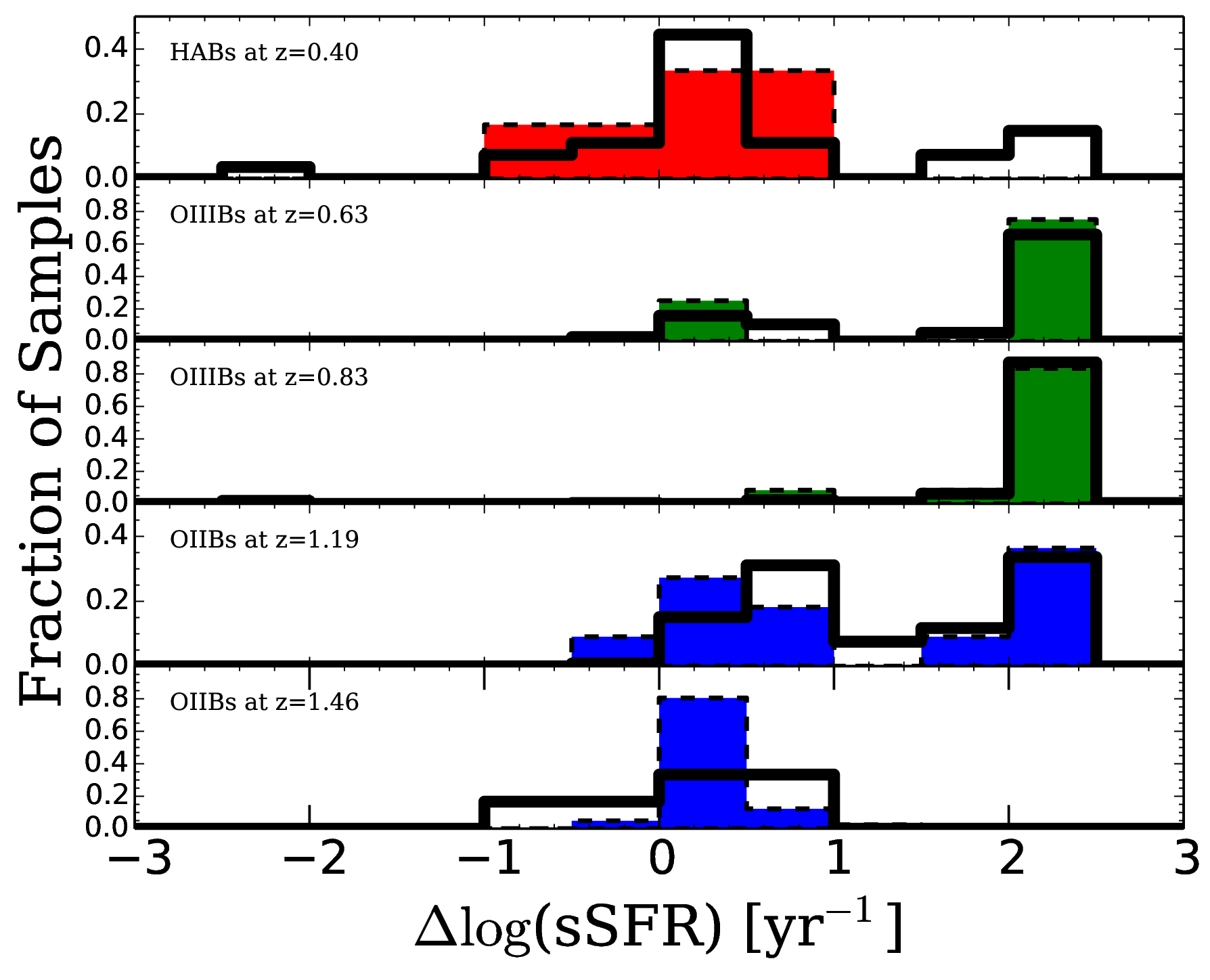}
	\caption{Normalized histograms of $\Delta$ sSFR compared to the main sequence at $z\sim1$ by 
	\cite{elbaz07} for blobs and emitters at $z=0.40-1.46$. 
	The meanings of symbols are identical to those in Figure \ref{fig_histogram_mass}. 
	}
	\label{fig_histogram_ssfr}
\end{figure}

Figure \ref{fig_histogram_mass} shows normalized histograms of the resulting stellar masses 
of the blobs from $z=0.40$ to $z=1.46$. 
For comparison purpose, we also plot the stellar-mass histograms of all emitters 
(including blob samples) in the figure. 
In order to make an unbiased comparison, 
we compare the SED fitting results of the blobs with those of the emitters 
that have the same emission-line fluxes (i.e., the same \nbest-\rz/\nbnto-\zp\ magnitude). 
In other words, the comparison is performed for the \ha, \oiii, \oii\ blobs/emitters with 
the corresponding line fluxes of $3-12\times10^{-16}$, $0.5-4\times10^{-16}$, and $0.4-5\times10^{-16}$ 
\ergscm, respectively. 
The histograms are normalized to show the fraction of blobs or emitters in each mass bin. 
It is seen at all redshifts that the blobs have obviously larger stellar masses 
than the emitters at the same redshift. 
At $z=0.40$, \ha\ emitters have the stellar masses ranging from $\sim5\times10^7$ \Msun\ to 
$\sim5\times10^{10}$ \Msun, while \hab s are located only at the massive end of the distribution. 
We perform the the Kolmogorov-Smirnov (KS) test to examine if the histograms of \hab s and \ha\ emitters 
are drawn from the same distribution. We can reject the null hypothesis 
that they are from the same distribution at the 90\% confidence level. 
Similarly, we can reject the null hypothesis for being drawn from the same distributions at the 99\% confidence level 
for \oiib s at $z=1.19$ a and at the 90\% confidence level for \oiib s at $z=1.46$. 
On the other hand, we cannot reject the null hypothesis that \oiiib s at $z=0.63$ and at $z=0.83$ 
are drawn from the same stellar mass distributions as the emitters at corresponding redshifts. 
In addition to the KS test, we perform the Anderson-Darling (AD) test, 
which is more sensitive than the Kolmogorov-Smirnov (KS) test, to 
examine the histograms. 
The results are almost consistent with the KS test, but at different confidence levels 
for \hab s at $z=0.40$ and \oiib s at $z=1.46$. With the AD test, the 95\% confidence level that 
we can reject the null hypothesis for \hab s at $z=0.40$ 
are higher than that obtained from the KS test, while the null hypothesis for \oiib s at $z=1.46$ 
can be rejected at the 75\% confidence level. 
For \oiib s at $z=1.19$, we can still reject the null hypothesis at the same 99\% confidence level. 
From both statistical tests, the blobs with spatially extended \ha\ or \oii\ emission lines 
are among those of the most massive emitters at the redshift. 
It is indicated that large-scale outflows are more prominent in the massive star-forming galaxies.

We plot the difference of specific star formation rate (sSFR) between our sample and 
the star formation main sequence at $z\sim1$ by \cite{elbaz07} in Figure \ref{fig_histogram_ssfr} 
to compare the star formation activity of blobs with those of the emitters 
with the same line fluxes. 
There are a number of \hab s with the sSFR lower than a majority of the \ha\ emitters ($\Delta \log$(sSFR) $<0.0$), 
suggesting that their star formation activity is relatively more quiescent than most 
of the star-forming galaxies at the same redshift. 
Interestingly, while most emitters are located around $\Delta \log$(sSFR) $=0$, 
a significant number of them show remarkably high sSFR, especially the \oiii\ emitters. 
This suggests that the \oiii\ emitters are mostly starbursts. 
At $z>1$, the histogram of \oii\ emitters is marginally shifted to $\Delta \log$(sSFR) greater than zero with 
30\% of \oii\ emitters at $z=1.19$ at the highest $\Delta \log$(sSFR). 
The blobs at each redshift have the same sSFR distributions as the emitters. 
According to the AD and the KS tests, we cannot rule out the null hypothesis that they are 
drawn from the same distribution. 

In conclusion, all types of blobs show a variety in star formation activity ranging from the 
most active star formation like a starburst 
to quiescent activity. 
It is implied that the physical mechanism driving the extended emission 
in the blobs can be individually different from each other. 
The only property that most blobs have in common is the stellar mass. 
\hab s and \oiib s are located at the massive end of the distributions 
of the corresponding normal emitters.

\begin{figure}
\centering
\begin{tabular}{ccc}
\includegraphics[width=0.29\textwidth]{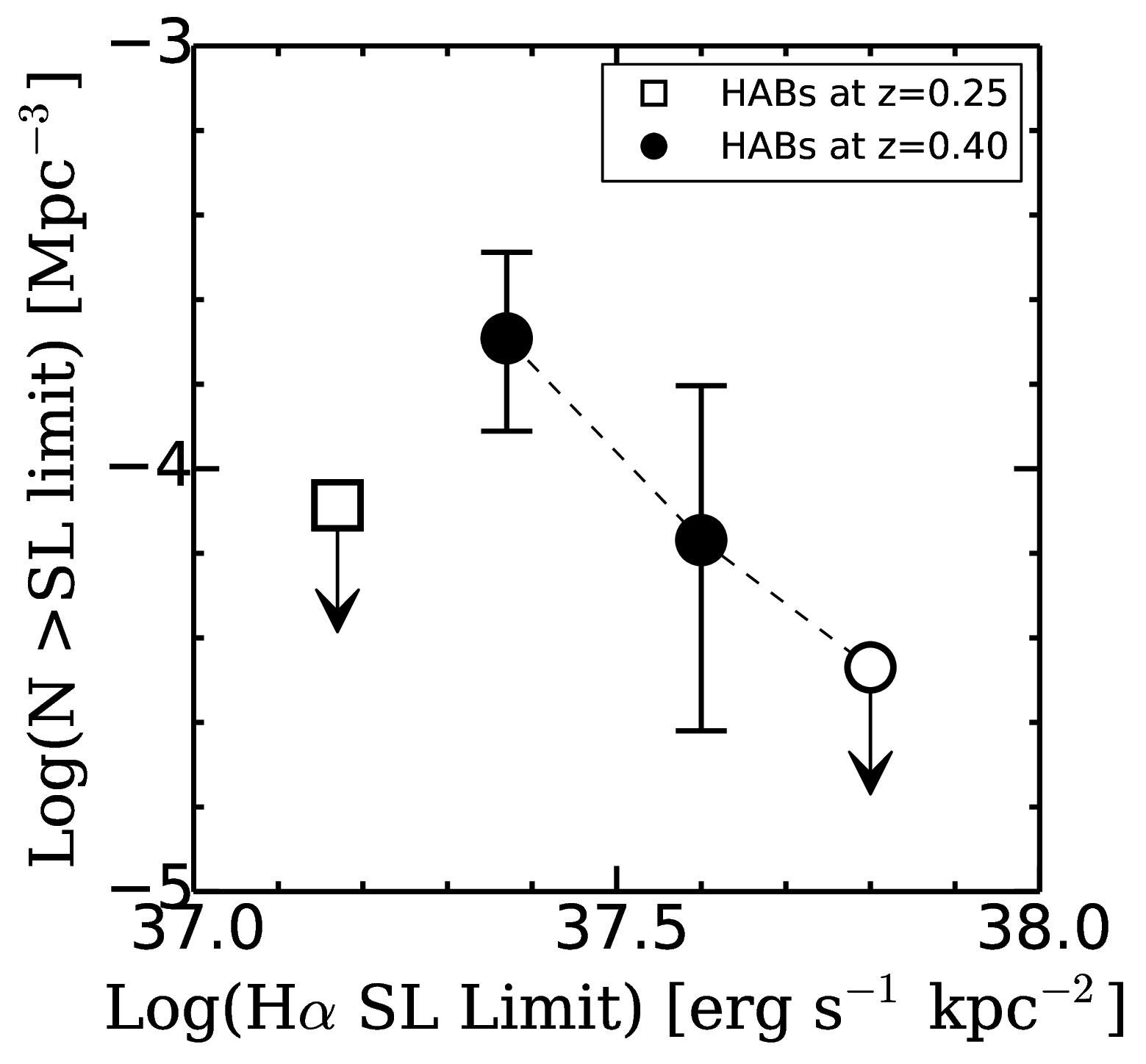} \\
\includegraphics[width=0.31\textwidth]{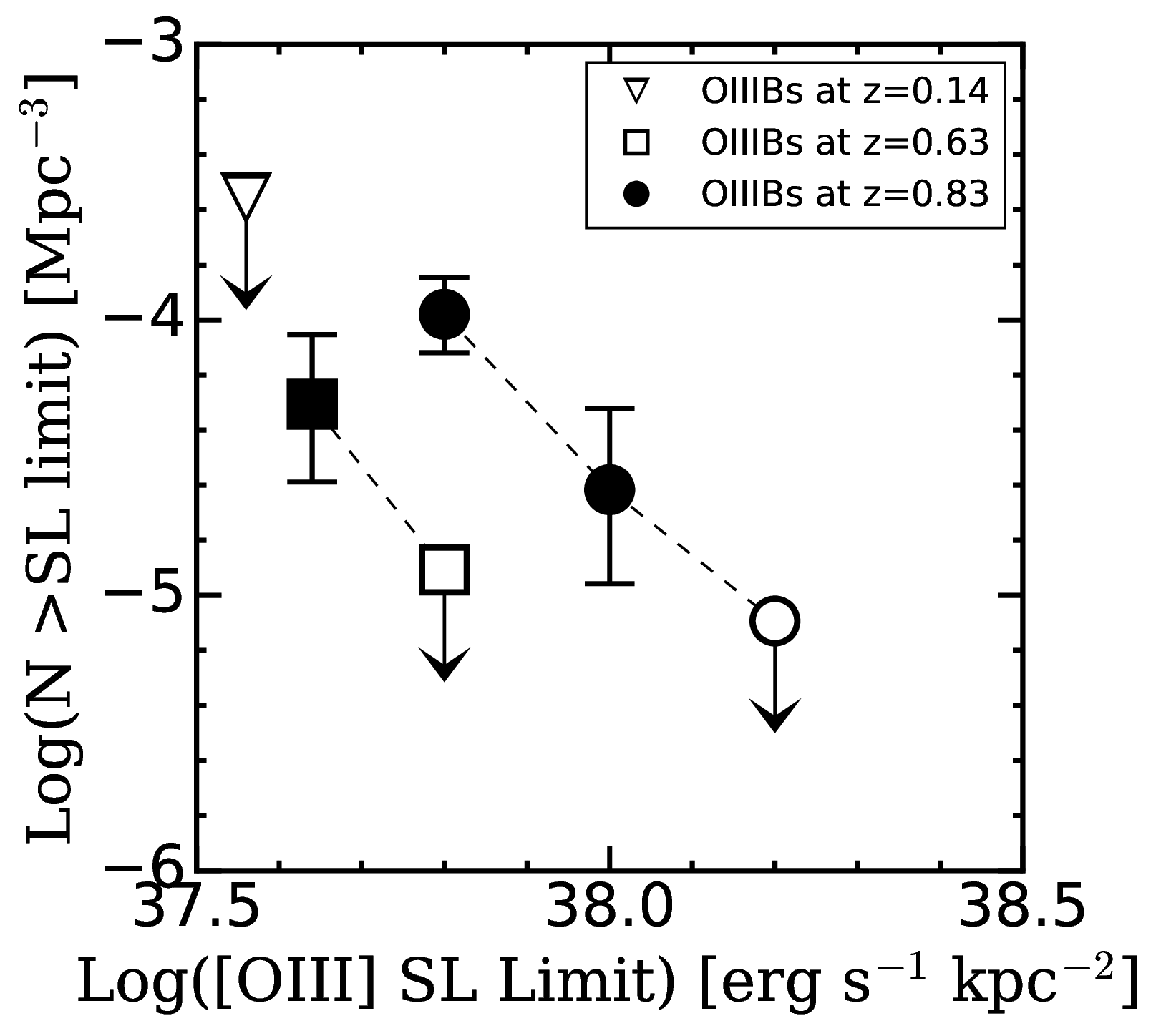} \\ 
\includegraphics[width=0.31\textwidth]{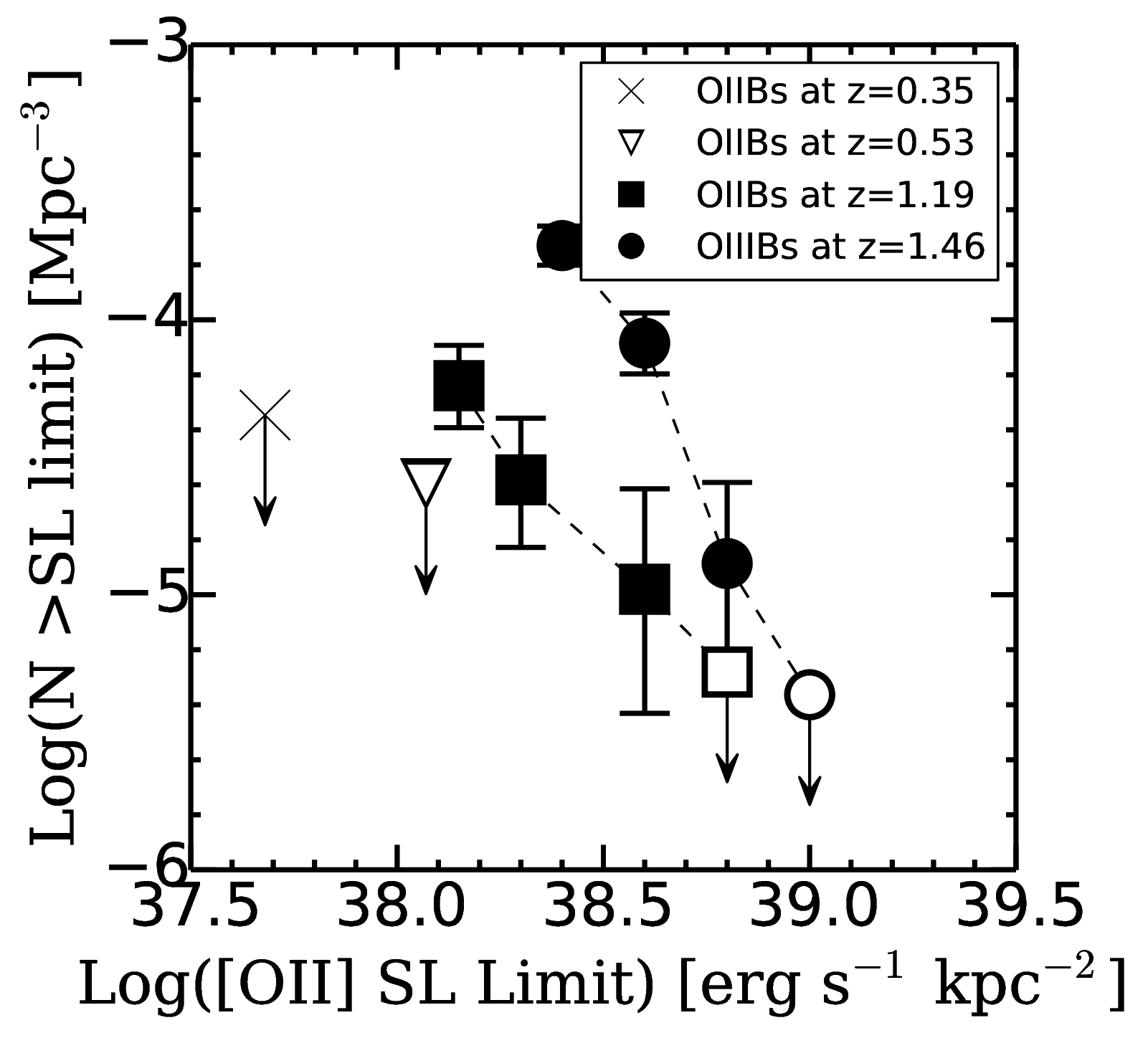} \\
\end{tabular}
\caption{Cumulative number density of the \hab s ({\it top}), \oiiib s ({\it middle}), 
and \oiib s ({\it bottom}) at $z=0.25-1.46$. 
All blobs are selected with isophotal area larger than 900 kpc$^2$ measured down to 
the surface luminosity indicated in the horizontal axis. 
Open symbols represent the upper limits, while circles, squares, triangles, 
and crosses indicate the number densities of blobs selected by using the 
\nbnto, \nbest, \nbfso, and \nbfot\ images, respectively. 
}
\label{fig_lf_sxds}
\end{figure}

\begin{figure}
	\centering
	\includegraphics[width=0.4\textwidth]{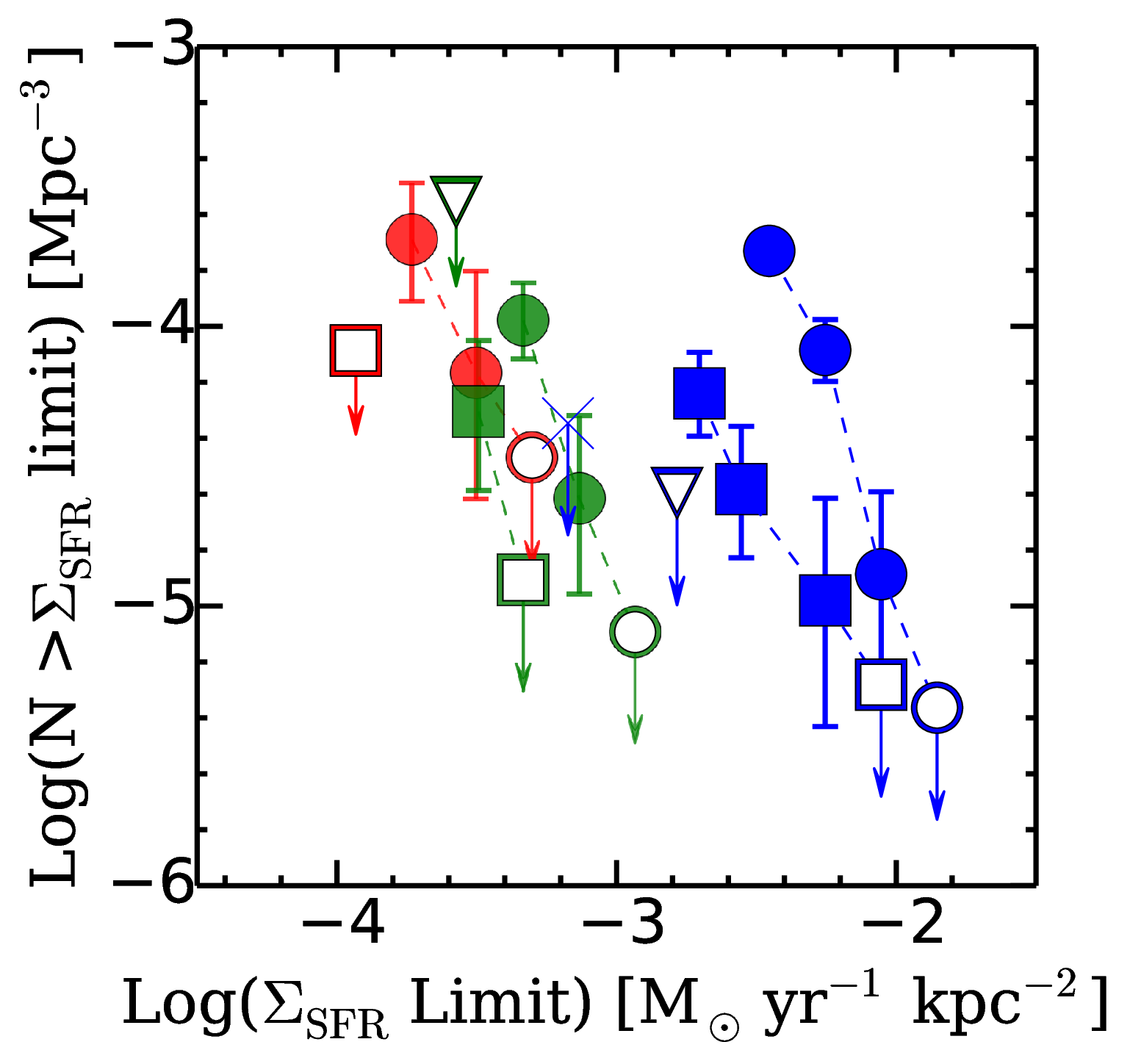}
	\caption{Cumulative number density  as a function of surface SFR limit for all types of blobs at $z=0.25-1.46$. 
		Similar to Figure \ref{fig_lf_sxds}, circles, squares, triangles, and crosses represent 
		blobs selected with \nbnto, \nbest, \nbfso, and \nbfot, respectively. 
		The upper limits of the number densities are shown with open symbols. 
		Red, green, and blue colors indicate the blobs with extended \ha, \oiii, and 
		\oii\ emission, respectively. 
	}
	\label{fig_numberden_sfr}
\end{figure}

\subsection{Cumulative Function of Surface Luminosity Limit}\label{subsec:lumi}

We plot the cumulative number densities of each type of blobs 
as a function of surface luminosity (SL) limits in Figure \ref{fig_lf_sxds}. 
The number density at each SL limit is derived 
by reselecting the blobs with isophotal area above 900 kpc$^2$ 
down to different SL limits. As a cumulative number density, 
the number density at the fainter SL limit, by definition, includes the number of blobs 
selected by the brighter limit. 
Therefore, the number density of blobs undoubtedly decreases 
with increasing the SL limit. 
Note that the SL limits shown in Figure \ref{fig_lf_sxds} are corrected 
for dust extinction by adopting the typical color excess of blobs ($E(B-V)=0.3$ mag) 
derived by the SED fitting 
and assuming the equivalent attenuation of the nebular and stellar components. 
This value is consistent with the extinction correction used in various studies 
of emitters \citep[e.g., ][]{drake13, sobral13}.

The top panel of Figure \ref{fig_lf_sxds} shows 
cumulative number densities of \hab s at $z=0.25$ and $z=0.40$. 
Although we have only an upper limit for the \hab s at $z=0.25$, 
the number density seems to be smaller than those at $z=0.4$ 
if we extrapolate the number density at $z=0.40$ to the SL limit for \hab s at $z=0.25$.  
(i.e., \ha\ SL limit $\sim10^{37.2}$ \ergs\,kpc$^{-2}$). 
Similarly, the number density of \oiiib s at $z=0.63$ is lower than 
the number density of the blobs at $z=0.83$ at the faintest end 
as seen in the middle panel of Figure \ref{fig_lf_sxds}. 
However, the upper limit of the number density of the blobs at $z=0.14$ is too high 
to make any conclusive discussion. 

The bottom panel of Figure \ref{fig_lf_sxds} shows 
the number densities of \oiib s at $z=0.35-1.46$. 
The number density decreases from $z=1.46$ to $z=1.19$ 
especially at the faint end of surface luminosity limits. 
Including the upper limits at $z=0.35$ and $z=0.53$, 
the number densities of \oiib s tend to decrease with redshifts. 
It is seen that the evolution of decreasing 
number density of blobs as a function of redshifts seems to be 
common in all types of blobs, though we only observe upper limits at some redshifts. 

\begin{figure}
		\centering
		\includegraphics[width=0.45\textwidth]{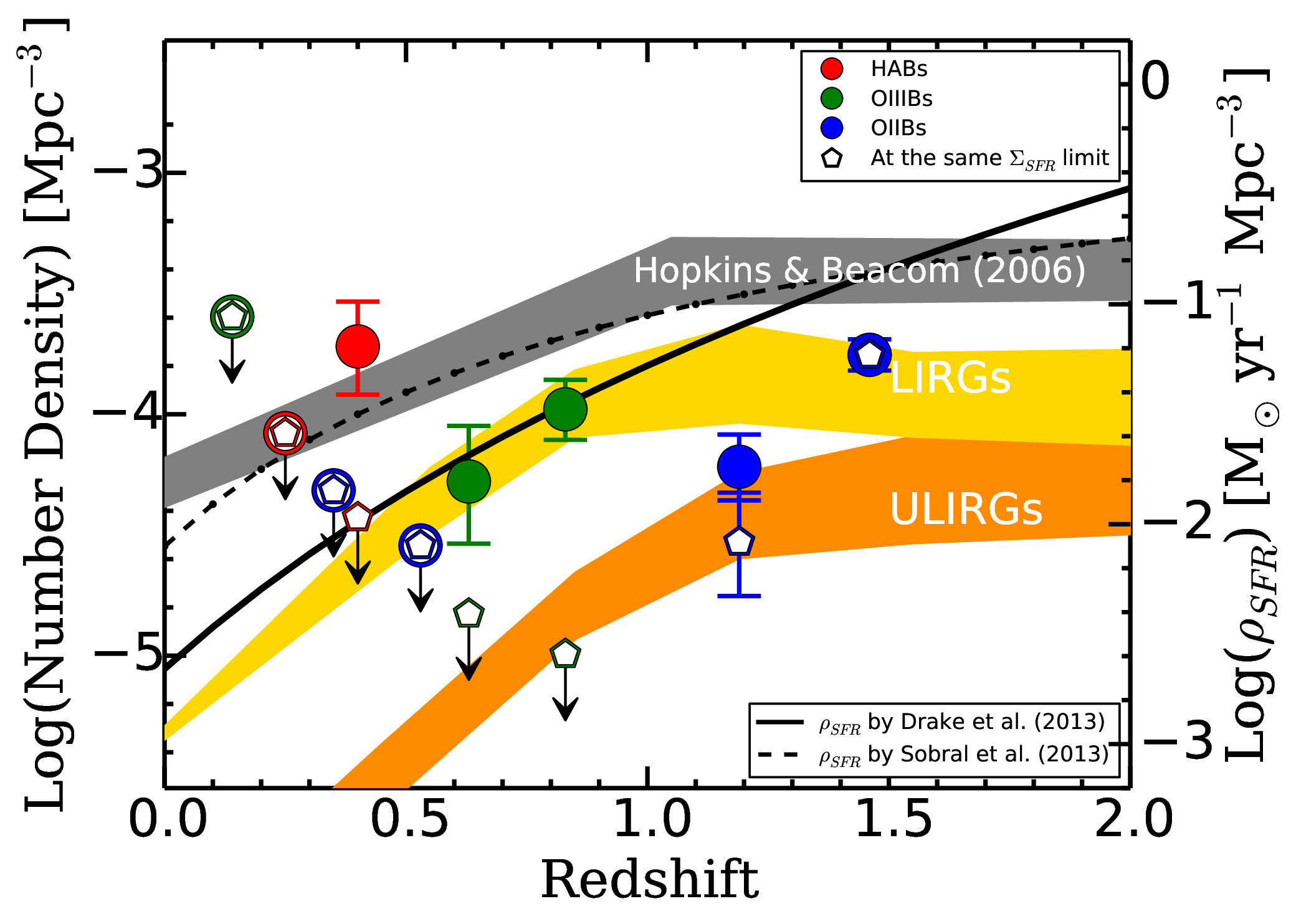}
		\caption{Number densities of \hab s (red), \oiiib s (green), and \oiib s (blue) at $z=0.14-1.46$ as a function of redshifts. 
			Circles and pentagons represent the blobs with isophotal area larger than 900 kpc$^2$ 
			measaured down to the 
			flux limit of $1.2\times10^{-18}$ \ergscm\ arcsec$^{-2}$ and the surface SFR limit 
			of $2.5\times10^{-3}$ \Msun yr$^{-1}$ kpc$^{-2}$, respectively. 
			Downward arrows indicate upper limits in the case of no selected blob. 
			Grey, yellow, and orange contours show the overall star formation rate densities of 
			star-forming galaxies \citep{hopkins06}, those of LIRGs, and ULIRGs \citep{casey14}, respectively. 
			Solid and dashed lines show the star formation rate densities of various emitters 
			at similar redshifts by \cite{drake13} and \cite{sobral13}, respectively. 
		}
		\label{fig_cosmic_lf}
\end{figure}

So far, we investigate the evolution of number densities separately for each 
type of blobs according to the emission lines we used to select them. 
We cannot directly compare blobs selected with different 
emission lines. 
In order to make a fair comparison across all types of blobs, 
we convert the surface limit of each emission-line luminosity into the surface SFR 
($\Sigma_{\rm SFR}$) limit. 
The \ha~and \oii~surface luminosities are converted to the surface SFR by 
using the relations given by \cite{kennicutt98}.  
For the \oiii\ emission line, we adopt the SFR calculation by \cite{drake13}, 
which uses the \ha/\hb\ ratio of 2.78 and \oiii/\hb\ ratio of 3.0 \citep{osterbrock06}. 
Although it is simple assumption for the star-forming galaxies, it is well 
in agreement with the typical line ratios of star-forming galaxies at $z=0.47-0.92$ 
\citep{nakajima13, lilly03}. 
Figure \ref{fig_numberden_sfr} shows the number densities of all blobs as a function of 
the $\Sigma_{\rm SFR}$ limit. 
It is seen that each type of blobs is located in different ranges of the $\Sigma_{\rm SFR}$ limit. 
This is solely due to the fact that each emission line relates to the SFR differently. 
The $\Sigma_{\rm SFR}$ limits for the \oiib s stand out of the others at high $\Sigma_{\rm SFR}$. 
Comparing within \oiib\ population, 
we find significant evolution that the number densities of the \oiib s tends to 
decrease with redshifts from $z=1.46$ to $z=1.19$. 
If we naively extrapolate the number densities of \oiib s at $z=1.19-1.46$ toward the 
lower $\Sigma_{\rm SFR}$ limit, 
the trend of declining number densities of blobs seems to continue down to $z=0.83$ 
(green circles in Figure \ref{fig_numberden_sfr}). 
At lower redshifts, the number densities show slight evolution from $z=0.83$ to $z=0.63$ 
and are comparable at $z<0.63$, 
suggesting no significant evolution in that period of the universe. 


\subsection{Number Density Evolutions of Blobs}\label{subsec:numberden}

In order to obtain a clearer picture of the evolution of blob number densities, 
we plot the number densities of all blobs selected down to the same flux limit 
as described in section \ref{sec:data} with filled circles in Figure \ref{fig_cosmic_lf}. 
The star formation rate densities (SFRD) of emitters \citep{drake13, sobral13}, 
star-forming galaxies \citep{hopkins06}, luminous infrared galaxies (LIRGs), 
and ultraluminous infrared galaxies \citep[ULIRGs; ][]{casey14} are shown to examine if 
the decrease in number density of blobs is similar to those of other galaxies. 
As discussed earlier, the number density tends to decrease with decreasing redshifts 
similar to the SFRD of the emitters, 
if we consider each type of blob separately. 
In contrast, the number densities seem to be almost constant within the error bars 
when comparing them across the types of blobs. 

In fact, the blobs that we select with the isophotal area down to the same flux limit cannot be 
compared across different types of emission lines at different redshifts. 
We thus re-select all types of blobs with the isophotal area larger than 900 kpc$^2$ down to 
the same $\Sigma_{\rm SFR}$ limit of $2.5\times10^{-3}$ \Msun yr$^{-1}$ kpc$^{-1}$ to 
make a fair comparison among all types of blobs at different redshifts. 
The resulting number densities are shown with open pentagons in Figure \ref{fig_cosmic_lf}. 
Unfortunately, we only obtain the number densities of blobs at the highest redshifts 
($z=1.19$ and $z=1.46$), 
while an upper limit is shown at $z < 1.0$. 
The number densities of blobs at $z\geq 0.83$ tend to decrease drastically toward lower redshifts. 
This decrease is even more rapid than the decline of cosmic star formation density, 
but it seems to be roughly comparable to the decline of ULIRGs. 
At $z<0.83$, the upper limits are not useful to interpret as the survey volume is probably 
too small to draw any conclusion. 
Larger and deeper survey is desirable to search for blobs with extended emission lines at low redshifts.

As mentioned in Section \ref{sec:intro}, we develop the method to select 
the blobs with spatially extended emission lines in order to systematically 
study the large-scale outflow. 
The outflow signature in a number of blobs is spectroscopically confirmed 
\citep[][Yuma et al. in prep.]{yuma13, harikane14}. 
If we naively consider all blobs as a galactic-scale outflow, we will be able to 
interpret the evolution of the blobs as that of the outflow events. 
Therefore, it is implied that the large-scale outflow events decrease significantly 
with redshifts at more rapid rate than the decline of the cosmic star formation density. 
The detailed study of the outflow process from blobs will be in the upcoming paper 
(Yuma et al. in prep.).

\section{Summary}\label{sec:summary}

Following the first systematic survey of \oii\ blobs at $z=1.19$ by Y13, 
we expand our search to galaxies with extended \oii $\lambda\lambda3726,3729$, 
\oiii $\lambda\lambda4959,5007$, and \ha $\lambda6563$ emission lines, which 
we call \oiib, \oiiib, and \hab, respectively.  
The redshifts of blob samples range from $z=0.14$ to $z=1.46$ covering 
roughly 9 Gyr of the Universe. 
We use all narrowband and broadband images available in the SXDS field; i.e., 
\nbfot, \nbfso, \nbest, \nbnto, B, V, R, \ip, and \zp. 
The blobs are selected with two isophotal area criteria: 
$>900$ kpc$^2$ and $>1500$ kpc$^2$. 
In this paper, we develop new quantitative approaches to 
effectively select the blobs with genuine extended emission lines. 
We simulate the bright PSFs and maesure their isophotal area 
to exclude the bright PSF-like emitters 
that can mimic the extended feature of the emission lines in the blobs. 
Furthermore, we create the equivalent-width map to 
examine if the extended feature of the emission line 
is real or it is just emission from the neighboring objects. 
This new method can efficiently exclude the emitters with 
very close nearby objects and is independent of personal bias, 
because we do not use visual inspection to determine the 
extended blobs as done by Y13. 
Our main results are summarized as follows: 
\begin{enumerate}
\item With the criterion of isophotal area larger than 900 kpc$^2$, 
we obtain 4 \oiiib s at $z=0.63$ and 11 \oiib s at $z=1.19$ selected 
with the \nbest\ image and 6 \hab s at $z=0.40$, 13 \oiiib s at $z=0.83$, 
and 43 \oiib s at $z=1.46$ selected with the \nbnto\ image. 

\item With the larger isophotal-area criterion ($>1500$ kpc$^2$), 
we find only one \hab\ at $z=0.40$, two \oiiib s at $z=0.83$, 
three \oiib s at $z=1.19$, and six \oiib s at $z=1.46$. 

\item We carry out the spectroscopic follow-up observations and 
confirm that two \oiiib s are at $z=0.62-0.64$ and the other two are at $z=0.84$. 
They all show significant spatial extension of the \oiii\ emission compared to their 
stellar components. 

\item We check the AGN contribution by using the X-ray and radio catalogs 
and found few percent of blobs are likely to host an AGN. The fraction of AGNs 
increases with isophotal area of the emission lines. 

\item We make a fair comparison of the stellar properties between 
the blobs and emitters with the same emission-line fluxes. 
Histograms of the stellar masses indicate that \hab s and \oiib s are on average 
more massive than the normal emitters at the same redshifts. 
However, we cannot reject the null hypothesis that the distributions of specific SFRs are 
drawn from the same distributions as the emitters.

\item As a systematic search, we are able to investigate the evolution of number densities of blobs 
and subsequently the evolution of the large-scale outflow events. 
Considering each type of the blob separately, we found 
a decreasing trend of number densities toward lower redshifts 
in \oiiib\ and \oiib\ samples with upper limits at some redshifts at the lowest end. 
However, if we compare the blobs across the different emission lines, we found no evolution at $z=0.40-0.63$. 
This comparison is slightly biased, because the same observed flux limit we used to measure the isophotal area 
corresponds to different luminosity limit at different redshifts. 

\item In order to make a fair comparison, we re-select the blobs down to 
the same surface star formation rate limit. 
Although the number densities become upper limits for most redshifts, 
we could at least constrain that the number densities of blobs 
drastically decrease toward low redshifts. 
The decreasing trend of the outflow events 
seems to be steeper than the decline 
of cosmic star formation density. 

\end{enumerate}

For most redshifts, we need deeper and larger imaging surveys to get rid of the upper limits 
and obtain the better constraint of the evolution of the blobs. 
The ongoing Hyper Suprime-Cam legacy survey with total 300 nights observed with the Subaru telescope will 
hopefully provide great number of candidates for blobs with extended emission lines, 
which are the galaxies with potentially large-scale outflows. 

\acknowledgements

The authors would like to thank an anonymous referee 
for detailed comments that improved clarity of this article. 
We also thank Christopher J. Conselice, the ApJ scientific editor, for useful comments 
on the Anderson-Darling test. 
This work is supported by Faculty of Science, Mahidol University, Thailand and 
the Thailand Research Fund (TRF) through research grant for new scholar (MRG5980153). 
S.Y thanks David J. Ruffolo for helpful suggestions as a mentor through the TRF research grant. 
S.Y. acknowledges support from the Japan Society for the Promotion of Science (JSPS) and 
the Talent Management program at Mahidol University, Thailand. 
M.O. is supported by 
World Premier International Research Center Initiative (WPI Initiative), MEXT, Japan, 
and KAKENHI (15H02064) Grant-in-Aid for Scientic Research (A) through JSPS. 
A.B.D. acknowledge ANR FOGHAR grant ANR-13-BS05-0010-02. 
This work is based on data collected at Subaru Telescope, which is operated by the 
National Astronomical Society of Japan.

\bibliographystyle{apj}
\bibliography{blob_evo_2017}

\appendix \label{appendix}
Catalogs of all blobs at $z=0.40-1.46$ are listed below. 
It is noteworthy that the observed flux and luminosity of the emission lines listed here 
are not corrected for the dust extinction yet. \\

\setcounter{table}{0}
\renewcommand{\thetable}{A\arabic{table}}

\begin{deluxetable*}{l l l cc cc}[!h]
\tabletypesize{\small}
\tablewidth{0pt}
\tablecolumns{11}
\tablecaption{Catalog of \hab s at $z=0.40$ in the SXDS field \label{tab:hab040}}
\tablewidth{0pt}
\tablehead{
\multicolumn{1}{c}{Object Name\tablenotemark{a}} &
\multicolumn{1}{c}{$\alpha$ (J2000)} &
\multicolumn{1}{c}{$\delta$ (J2000)} &
\multicolumn{1}{c}{\nbnto-\zp\tablenotemark{b}} & 
\multicolumn{1}{c}{$F$(\ha)} & 
\multicolumn{1}{c}{$L$(\ha)} & 
\multicolumn{1}{c}{Isophotal Area} \\
\multicolumn{1}{c}{} &
\multicolumn{1}{c}{} &
\multicolumn{1}{c}{} &
\multicolumn{1}{c}{(mag)} & 
\multicolumn{1}{c}{($10^{-16}$ \ergscm)} & 
\multicolumn{1}{c}{($10^{41}$ \ergs)} & 
\multicolumn{1}{c}{(arcsec$^2$)} 
}

\startdata
\multicolumn{7}{c}{Isophotal area more than 1500 kpc$^2$}\\
\midrule
HAB040m-1 & 02 16 11.787 & -04 45 16.952 & $20.39\pm0.09$ & $11.8\pm1.1$& $6.7\pm0.6$& 54\\
\midrule
\multicolumn{7}{c}{Isophotal area more than 900 kpc$^2$}\\
\midrule
HAB040s-2 &  02 17 21.392 & -04 43 02.577 & $20.78\pm0.12$ & $8.3\pm1.0$ & $4.7\pm0.6$& 45\\
HAB040m-3 & 02 17 24.184 & -05 25 26.949 & $20.67\pm0.11$ & $9.2\pm1.1$ & $5.2\pm0.6$& 40\\
HAB040m-4 & 02 17 23.373 & -05 13 53.316 & $21.34\pm0.13$ & $5.0\pm0.6$ & $2.8\pm0.3$& 37\\
HAB040m-5 & 02 17 03.259 & -05 24 23.957 & $20.67\pm0.12$ & $9.1\pm1.1$ & $5.2\pm0.6$& 35\\
HAB040m-6 & 02 18 59.135 & -05 23 37.375 & $21.24\pm0.14$ & $5.4\pm0.8$ & $3.1\pm0.4$& 33
\enddata
\tablenotetext{a}{The letters "m" and "s" after HAB040 mean "multiple" and "single", respectively. They are used to indicate the number of continuum components of the blobs.}
\tablenotetext{b}{\nbnto-\zp\ magnitudes are measured down to $1.2\times 10^{-18}$ \ergscm\, arcsec$^{-2}$.}
\end{deluxetable*}

\begin{deluxetable*}{l ll cc cc}[!h]
\tabletypesize{\small}
\tablewidth{0pt}
\tablecolumns{11}
\tablecaption{Catalog of \oiiib s at $z=0.63$ in the SXDS field\label{tab:o3b063}}
\tablewidth{0pt}
\tablehead{
\multicolumn{1}{c}{Object Name} &
\multicolumn{1}{c}{$\alpha$ (J2000)} &
\multicolumn{1}{c}{$\delta$ (J2000)} &
\multicolumn{1}{c}{\nbest-\rz\tablenotemark{a}} & 
\multicolumn{1}{c}{$F$(\oiii)} & 
\multicolumn{1}{c}{$L$(\oiii)} & 
\multicolumn{1}{c}{Isophotal Area} \\
\multicolumn{1}{c}{} &
\multicolumn{1}{c}{} &
\multicolumn{1}{c}{} &
\multicolumn{1}{c}{(mag)} & 
\multicolumn{1}{c}{($10^{-16}$ \ergscm)} & 
\multicolumn{1}{c}{($10^{41}$ \ergs)} & 
\multicolumn{1}{c}{(arcsec$^2$)} 
}

\startdata
\multicolumn{7}{c}{Isophotal area more than 900 kpc$^2$}\\
\midrule

OIIIB063s-1 & 02 17 50.244& -05 00 04.159 & $22.04\pm0.14$ & $3.0\pm0.4$ & $4.99\pm0.7$ & 24 \\
OIIIB063s-2 & 02 19 05.902& -05 13 48.599 & $22.27\pm0.15$ & $2.4\pm0.4$ & $4.04\pm0.7$ & 24 \\
OIIIB063s-3 & 02 18 31.564& -05 24 24.611 & $22.12\pm0.15$ & $2.8\pm0.4$ & $4.65\pm0.7$ & 23 \\
OIIIB063s-4 & 02 17 09.652& -04 44 16.176 & $21.71\pm0.15$ & $4.0\pm0.6$ & $6.76\pm1.0$ & 22 
\enddata
\tablenotetext{}{{\bf Note:} There is no \oiiib\ at $z=0.63$ with the 
isophotal area larger than 1500 kpc$^2$ and all \oiiib s at this redshift have only one stellar component. }
\tablenotetext{a}{\nbest-\rz\ magnitudes are measured down to $1.2\times 10^{-18}$ \ergscm\, arcsec$^{-2}$.}
\end{deluxetable*}
\begin{deluxetable*}{l l l  cc cc}[!h]
\tabletypesize{\small}
\tablewidth{0pt}
\tablecolumns{11}
\tablecaption{Catalog of \oiiib s at $z=0.83$ in the SXDS field\label{tab:o3b083}}
\tablewidth{0pt}
\tablehead{
\multicolumn{1}{c}{Object Name\tablenotemark{a}} &
\multicolumn{1}{c}{$\alpha$ (J2000)} &
\multicolumn{1}{c}{$\delta$ (J2000)} &
\multicolumn{1}{c}{\nbnto-\zp\tablenotemark{b}} & 
\multicolumn{1}{c}{$F$(\oiii)} & 
\multicolumn{1}{c}{$L$(\oiii)} & 
\multicolumn{1}{c}{Isophotal Area} \\
\multicolumn{1}{c}{} &
\multicolumn{1}{c}{} &
\multicolumn{1}{c}{} &
\multicolumn{1}{c}{(mag)} & 
\multicolumn{1}{c}{($10^{-16}$ \ergscm)} & 
\multicolumn{1}{c}{($10^{41}$ \ergs)} & 
\multicolumn{1}{c}{(arcsec$^2$)} 
}

\startdata
\multicolumn{7}{c}{Isophotal area more than 1500 kpc$^2$}\\
\midrule
OIIIB083m-1 & 02 18 58.757 & -05 16 57.162 & $21.89\pm0.11$ & $3.0\pm0.3$& $9.8\pm1.0$ & 35\\
OIIIB083m-2 & 02 19 03.734 & -05 11 53.243 & $22.77\pm0.13$ & $1.3\pm0.2$ & $4.4\pm0.6$ & 29\\
\midrule
\multicolumn{7}{c}{Isophotal area more than 900 kpc$^2$}\\
\midrule
OIIIB083m-3 & 02 19 03.780 & -05 28 53.328  & $22.05\pm0.18$ & $2.6\pm0.5$ & $8.48\pm1.5$ & 21\\
OIIIB083m-4 & 02 19 06.479 & -05 10 57.034  & $23.55\pm0.21$ & $0.6\pm0.1$& $2.14\pm0.3$ & 20\\
OIIIB083m-5 & 02 18 13.016 & -04 48 48.277  & $23.52\pm0.22$ & $0.7\pm0.2$& $2.18\pm0.6$ & 19\\
OIIIB083s-6 & 02 18 05.812 & -04 40 15.911   & $22.44\pm0.21$ & $1.8\pm0.4$ & $5.94\pm1.2$ & 19\\
OIIIB083m-7 & 02 18 27.313 & -05 17 42.059  & $22.08\pm0.20$ & $2.5\pm0.5$ & $8.25\pm1.5$ & 19\\
OIIIB083m-8 & 02 18 47.713 & -05 24 35.981  & $22.59\pm0.23$ & $1.6\pm0.4$ & $5.15\pm1.2$ & 19\\
OIIIB083s-9 & 02 18 54.318 & -04 47 34.030   & $22.49\pm0.22$ & $1.7\pm0.4$ & $5.68\pm1.2$ & 18\\
OIIIB083s-10 & 02 17 46.527 & -04 44 13.844 & $23.20\pm0.23$ & $0.9\pm0.2$ & $2.94\pm0.6$ & 18\\
OIIIB083m-11 & 02 17 56.121 & -04 46 09.748 & $21.95\pm0.22$ & $2.8\pm0.6$ & $9.29\pm1.8$ & 17\\
OIIIB083m-12 & 02 18 58.701 & -05 12 58.126 & $23.32\pm0.23$ & $0.8\pm0.2$ & $2.64\pm0.6$ & 17\\
OIIIB083s-13 & 02 19 03.915 & -04 46 57.229  & $23.16\pm0.25$ & $0.9\pm0.2$ & $3.06\pm0.6$ & 16
\enddata
\tablenotetext{a}{The letters "m" and "s" after OIIIB083 mean "multiple" and "single", respectively. They are used to indicate the number of continuum components of the blobs.}
\tablenotetext{b}{\nbnto-\zp\ magnitudes are measured down to $1.2\times 10^{-18}$ \ergscm\, arcsec$^{-2}$.}
\end{deluxetable*}

\begin{deluxetable*}{l ll cc cc}[!h]
\tabletypesize{\small}
\tablewidth{0pt}
\tablecolumns{11}
\tablecaption{Catalog of \oiib s at $z=1.19$ in the SXDS field\label{tab:o2b119}}
\tablewidth{0pt}
\tablehead{
\multicolumn{1}{c}{Object Name\tablenotemark{a}} &
\multicolumn{1}{c}{$\alpha$ (J2000)} &
\multicolumn{1}{c}{$\delta$ (J2000)} &
\multicolumn{1}{c}{\nbest-\rz\tablenotemark{b}} & 
\multicolumn{1}{c}{$F$(\oii)} & 
\multicolumn{1}{c}{$L$(\oii)} & 
\multicolumn{1}{c}{Isophotal Area} \\
\multicolumn{1}{c}{} &
\multicolumn{1}{c}{} &
\multicolumn{1}{c}{} &
\multicolumn{1}{c}{(mag)} & 
\multicolumn{1}{c}{($10^{-16}$ \ergscm)} & 
\multicolumn{1}{c}{($10^{41}$ \ergs)} & 
\multicolumn{1}{c}{(arcsec$^2$)} 
}

\startdata
\multicolumn{7}{c}{Isophotal area more than 1500 kpc$^2$}\\
\midrule
OIIB119m-1 & 02 17  08.642 & -04 50 22.876 & $22.01\pm0.10$ & $3.1\pm0.3$ & $24.6\pm2.4$& 34\\
OIIB119m-2 & 02 17  16.345 & -04 57 11.938 & $21.78\pm0.11$ & $3.8\pm0.4$ & $3.03\pm3.2$& 31\\
OIIB119m-3 & 02 18  55.226 & -05 21 17.681 & $23.30\pm0.15$ & $0.9\pm0.1$ & $7.5\pm0.8$& 26\\

\midrule
\multicolumn{7}{c}{Isophotal area more than 900 kpc$^2$}\\
\midrule
OIIB119m-4 & 02 19  19.090 & -05 15 35.280 	& $22.77\pm0.17$ & $1.5\pm0.2$& $12.2\pm1.6$& 21\\
OIIB119m-5 & 02 17  34.407 & -04 58 59.858 	& $23.40\pm0.18$ & $0.9\pm0.2$& $6.8\pm1.6$& 20\\
OIIB119s-6 & 02 19  01.684 & -04 59 33.233 	& $23.68\pm0.21$ & $0.7\pm0.1$& $5.3\pm0.8$& 18\\
OIIB119s-7 & 02 18  56.110 & -04 45 50.865 	& $23.23\pm0.20$ & $0.9\pm0.2$& $8.0\pm1.6$& 17\\
OIIB119s-8 & 02 17  45.842 & -04 42 40.343 	& $23.16\pm0.20$ & $1.1\pm0.2$& $8.5\pm1,6$& 17\\
OIIB119m-9 & 02 18  59.747 & -04 49 41.717 	& $23.68\pm0.24$ & $0.7\pm0.2$& $5.3\pm1,6$& 16\\
OIIB119s-10 & 02 18  53.660 & -05 15 23.773 	& $23.87\pm0.26$ & $0.6\pm0.2$& $4.5\pm1.6$& 14\\
OIIB119s-11 & 02 17  32.526 & -04 57 46.401 	& $22.90\pm0.25$ & $1.4\pm0.3$& $10.8\pm2.4$& 14

\enddata
\tablenotetext{a}{The letters "m" and "s" after OIIB119 mean "multiple" and "single", respectively. They are used to indicate the number of continuum components of the blobs.}
\tablenotetext{b}{\nbest-\rz\ magnitudes are measured down to $1.2\times 10^{-18}$ \ergscm\, arcsec$^{-2}$.}
\end{deluxetable*}
\begin{deluxetable*}{l ll cc cc}[!h]
\tabletypesize{\small}
\tablewidth{0pt}
\tablecolumns{11}
\tablecaption{Catalog of \oiib s at $z=1.46$ in the SXDS field\label{tab:o2b146}}
\tablewidth{0pt}
\tablehead{
\multicolumn{1}{c}{Object Name\tablenotemark{a}} &
\multicolumn{1}{c}{$\alpha$ (J2000)} &
\multicolumn{1}{c}{$\delta$ (J2000)} &
\multicolumn{1}{c}{\nbnto-\zp\tablenotemark{b}} & 
\multicolumn{1}{c}{$F$(\oii)} & 
\multicolumn{1}{c}{$L$(\oii)} & 
\multicolumn{1}{c}{Isophotal Area} \\
\multicolumn{1}{c}{} &
\multicolumn{1}{c}{} &
\multicolumn{1}{c}{} &
\multicolumn{1}{c}{(mag)} & 
\multicolumn{1}{c}{($10^{-16}$ \ergscm)} & 
\multicolumn{1}{c}{($10^{41}$ \ergs)} & 
\multicolumn{1}{c}{(arcsec$^2$)} 
}

\startdata
\multicolumn{7}{c}{Isophotal area more than 1500 kpc$^2$}\\
\midrule
OIIB146m-1 & 02 19 6.034 & -05 03 35.552 	& $22.50\pm0.13$ & $1.7\pm0.2$& $22.6\pm2.7$ & 36\\
OIIB146s-2 & 02 16 50.983 & -05 01 37.451 	& $22.64\pm0.13$ & $1.5\pm0.2$& $19.9\pm2.7$ & 28\\
OIIB146m-3 & 02 17 04.783 & -05 15 18.450 	& $22.72\pm0.17$ & $1.4\pm0.2$& $18.5\pm2.7$ & 28\\
OIIB146m-4 & 02 19 06.426 & -04 57 55.386 	& $22.45\pm0.15$ & $1.8\pm0.3$& $23.8\pm4.0$ & 26\\
OIIB146m-5 & 02 17 47.428 & -05 12 59.786 	& $22.96\pm0.16$ & $1.1\pm0.2$& $14.9\pm2.7$ & 25\\
OIIB146m-6 & 02 17 00.352 & -05 01 50.586 	& $22.94\pm0.19$ & $1.1\pm0.2$ & $15.1\pm2.7$ & 22\\

\midrule
\multicolumn{7}{c}{Isophotal area more than 900 kpc$^2$}\\
\midrule
OIIB146s-7 & 02 19 04.834 & -04 54 14.670 	& $23.45\pm0.20$& $0.7\pm0.1$& $9.45\pm1.3$ & 21\\
OIIB146s-8 & 02 19 02.692 & -04 47 23.322 	& $23.01\pm0.19$& $1.06\pm0.2$ & $14.1\pm2.7$ & 21\\
OIIB146m-9 & 02 17 23.563 & -05 05 40.962 	& $23.26\pm0.20$& $0.8\pm0.2$& $11.3\pm2.7$ & 21\\
OIIB146m-10 & 02 18 26.308 & -05 12 57.343 	& $23.06\pm0.19$& $1.01\pm0.2$ & $13.5\pm2.7$ & 21\\
OIIB146m-11 & 02 18 26.206 & -05 12 56.843 	& $23.06\pm0.19$& $1.01\pm0.2$ & $13.5\pm2.7$ & 21\\
OIIB146s-12 & 02 18 12.194 & -05 01 11.636 	& $22.45\pm0.19$& $1.77\pm0.3$ & $23.6\pm4.0$ & 20\\
OIIB146m-13 & 02 18 50.247 & -05 00 46.797 	& $22.64\pm0.21$& $1.50\pm0.3$ & $19.9\pm4.0$ & 20\\
OIIB146m-14 & 02 17 13.236 & -05 21 41.889 	& $23.30\pm0.21$& $0.8\pm0.2$& $10.8e\pm2.7$+42& 20\\
OIIB146m-15 & 02 17 13.626 & -05 09 39.486 	& $23.21\pm0.21$& $0.8\pm0.2$& $11.8\pm2.7$ & 20\\
OIIB146s-16 & 02 16 58.166 & -05 13 39.015 	& $22.68\pm0.22$& $1.44\pm0.3$ & $19.2\pm4.0$ & 20\\
OIIB146m-17 & 02 19 18.176 & -04 47 31.143 	& $22.10\pm0.19$& $2.45\pm0.4$ & $32.7\pm5.3$ & 20\\
OIIB146s-18 & 02 19 17.159 & -04 52 10.298 	& $22.73\pm0.21$& $1.38\pm0.3$ & $18.3\pm4.0$ & 19\\
OIIB146m-19 & 02 17 34.183 & -05 10 16.175 	& $22.22\pm0.20$& $2.19\pm0.4$ & $29.2\pm5.3$ & 19\\
OIIB146s-20 & 02 17 49.028 & -05 21 38.785 	& $23.32\pm0.21$& $0.8\pm0.2$& $10.7\pm2.7$ & 19\\
OIIB146m-21 & 02 17 38.914 & -05 12 36.599 	& $22.86\pm0.22$& $1.23\pm0.3$ & $16.3\pm4.0$ & 19\\
OIIB146m-22 & 02 18 17.559 & -05 03 58.961 	& $22.51\pm0.21$& $1.69\pm0.3$ & $22.5\pm4.0$ & 19\\
OIIB146m-23 & 02 18 53.618 & -05 12 13.477 	& $22.60\pm0.21$& $1.55\pm0.3$ & $20.7\pm4.0$ & 19\\
OIIB146s-24 & 02 18 58.056 & -05 21 37.365 	& $23.26\pm0.21$& $0.8\pm0.2$& $11.3\pm2.7$ & 19\\
OIIB146m-25 & 02 17 09.097 & -04 50 47.034 	& $22.76\pm0.23$& $1.34\pm0.3$ & $17.8\pm4.0$ & 19\\
OIIB146m-26 & 02 17 05.657 & -04 55 47.036 	& $22.61\pm0.21$& $1.54\pm0.3$ & $20.5\pm4.0$ & 18\\
OIIB146m-27 & 02 17 12.991 & -04 54 40.741 	& $22.47\pm0.21$& $1.76\pm0.4$ & $23.4\pm5.3$ & 18\\
OIIB146m-28 & 02 18 56.884 & -05 00 56.060 	& $23.82\pm0.25$& $0.5\pm0.1$& $6.70\pm1.3$ & 18\\
OIIB146m-29 & 02 17 07.843 & -04 50 00.902 	& $22.71\pm0.23$& $1.40\pm0.3$ & $18.6\pm4.0$ & 18\\
OIIB146s-30 & 02 17 01.233 & -05 28 43.889 	& $23.56\pm0.24$& $0.6\pm0.1$& $8.54\pm1.3$ & 17\\
OIIB146m-31 & 02 18 26.263 & -04 40 55.116 	& $22.81\pm0.23$& $1.27\pm0.3$ & $17.0\pm4.0$ & 17\\
OIIB146m-32 & 02 17 58.677 & -05 12 42.674 	& $23.41\pm0.24$& $0.7\pm0.2$& $9.84\pm2.7$ & 17\\
OIIB146m-33 & 02 17 28.558 & -05 13 45.305 	& $22.56\pm0.24$& $1.61\pm0.4$ & $21.4\pm5.3$ & 16\\
OIIB146m-34 & 02 17 13.757 & -05 00 01.130 	& $22.74\pm0.25$& $1.36\pm0.3$ & $18.1\pm4.0$ & 16\\
OIIB146m-35 & 02 17 20.349 & -04 43 53.291 	& $23.40\pm0.25$& $0.7\pm0.2$& $9.85\pm2.7$ & 16\\
OIIB146m-36 & 02 17 12.632 & -04 54 43.034 	& $23.42\pm0.25$& $0.7\pm0.2$& $9.71\pm2.7$ & 16\\
OIIB146m-37 & 02 17 14.949 & -05 05 28.282 	& $23.09\pm0.25$& $1.0\pm0.2$& $13.1\pm2.7$ & 15\\
OIIB146s-38 & 02 19 12.393 & -05 13 17.377 	& $23.47\pm0.27$& $0.7\pm0.2$& $9.30\pm2.7$ & 15\\
OIIB146m-39 & 02 17 47.709 & -05 23 07.385 	& $23.34\pm0.28$& $0.7\pm0.2$& $10.5\pm2.7$ & 14\\
OIIB146s-40 & 02 18 30.241 & -05 04 10.890 	& $22.97\pm0.28$& $1.10\pm0.3$ & $14.7\pm4.0$ & 14\\
OIIB146m-41 & 02 19 04.278 & -05 19 59.379 	& $23.61\pm0.31$& $0.6\pm0.2$& $8.17\pm2.7$ & 13\\
OIIB146m-42 & 02 18 34.744 & -05 03 00.816 	& $22.73\pm0.30$& $1.38\pm0.4$ & $18.4\pm5.3$ & 13\\
OIIB146m-43 & 02 17 47.512 & -05 10 53.431 	& $23.42\pm0.29$& $0.7\pm0.2$& $9.67\pm2.7 $ & 13

\enddata
\tablenotetext{a}{The letters "m" and "s" after OIIB146 mean "multiple" and "single", respectively. They are used to indicate the number of continuum components of the blobs.}
\tablenotetext{b}{\nbnto-\zp\ magnitudes are measured down to $1.2\times 10^{-18}$ \ergscm\, arcsec$^{-2}$.}
\end{deluxetable*}

\end{document}